\documentclass[aps,prd,10pt,twocolumn,superscriptaddress,nofootinbib,noshowpacs,preprintnumbers,hyperref,floatfix,reprint]{revtex4-1}

\pdfoutput=1

\usepackage{amsmath}
\usepackage[utf8]{inputenc}
\usepackage{multirow}
\usepackage{graphicx}
\usepackage{xfrac}
\usepackage[colorlinks]{hyperref}
\hypersetup{allcolors=[rgb]{0,0,1}}
\usepackage[dvipsnames]{xcolor} 
\usepackage{lipsum, babel}

\usepackage{tikz,xcolor,hyperref}

\definecolor{lime}{HTML}{A6CE39}
\DeclareRobustCommand{\orcidicon}{\hspace{-1mm}
	\begin{tikzpicture}
	\draw[lime, fill=lime] (0,0) 
	circle [radius=0.16] 
	node[white] {{\fontfamily{qag}\selectfont \tiny \,ID}};
	\draw[white, fill=white] (-0.0525,0.095) 
	circle [radius=0.007];
	\end{tikzpicture}
	\hspace{-3mm}
}

\foreach \x in {A, ..., Z}{\expandafter\xdef\csname orcid\x\endcsname{\noexpand\href{https://orcid.org/\csname orcidauthor\x\endcsname}
			{\noexpand\orcidicon}}
}

\begin{document}

\title{Standing Accretion Shock Instability in the Collapse of a Rotating Stellar Core}

\author{Laurie Walk\orcidA{}}
\affiliation{Niels Bohr International Academy and DARK, Niels Bohr Institute, University of Copenhagen, Blegdamsvej 17, 2100, Copenhagen, Denmark}

\author{Thierry Foglizzo\orcidB{}}
\email{foglizzo@cea.fr}
\affiliation{Universit\'e Paris-Saclay, Universit\'e Paris Cit\'e, CEA, CNRS, AIM, 91191, Gif-sur-Yvette, France}

\author{Irene Tamborra\orcidC{}}
\email{tamborra@nbi.ku.dk}
\affiliation{Niels Bohr International Academy and DARK, Niels Bohr Institute, University of Copenhagen, Blegdamsvej 17, 2100, Copenhagen, Denmark}

\date{\today}

\begin{abstract}
Hydrodynamical instabilities, such as the  standing accretion shock instability (SASI), play an essential role in the dynamics of core-collapse supernovae, with observable imprints in the neutrino and gravitational wave signals. Yet, the impact of stellar rotation on the development of SASI is  poorly explored. 
We investigate the conditions favoring the growth of SASI in the presence of rotation  through a perturbative analysis. The properties of SASI are compared in two stationary configurations, cylindrical and spherical equatorial, which mainly differ by their advection timescales from the shock to the proto-neutron star surface. Without rotation,  the mode $m=1$, corresponding to a one-armed spiral SASI deformation, can be significantly more unstable in the spherical equatorial configuration.  In fact, the shorter advection time in the spherical equatorial geometry allows for a larger contribution of the entropic-acoustic coupling from the region of adiabatic compression near the surface of the proto-neutron star.
The angular momentum of the collapsing core favors the growth of prograde spiral modes $m=1$ and $m = 2$ in both geometries.  Although the growth rate of the spiral instability is systematically faster in spherical geometry,  its oscillation frequency is remarkably insensitive to the geometry. Such a contrast with non-rotating flows calls for a deeper understanding of the role of advection in the mechanism of spiral SASI. 
Our findings suggest that the resonant coupling of acoustic waves with their corotation radius may play a major role in the instability mechanism of collapsing cores with rotation. Elucidating this physical mechanism is essential to interpret the signal from future multi-messenger supernova observations.
\end{abstract}

\maketitle

\section{Introduction} 
The impact of angular momentum on the mechanism of core-collapse supernovae (SNe) is still incomplete~\cite{Janka:2017vcp,Burrows:2020qrp,Muller:2020}. Rotation may affect the structure of the pre-SN star, including non-negligible angular momentum accreted during the  core collapse of a massive star. It can amplify  the magnetic field with consequences on the  explosion mechanism and eventually play a key role in the magnetorotational mechanism. Rotation can also affect the production of the $r$-process elements, as well as the pulsar spin period~\cite{Heger:2004qp,Ott:2005wh,Blondin:2007,Winteler:2012hu,Mosta:2014jaa,Nishimura:2015nca,Obergaulinger:2017qno,Obergaulinger:2018udn}. The advent of  SN modeling in multi dimensions  has highlighted the role of rotation in  facilitating the explosion, see e.g.~\cite{Summa:2017wxq,Takiwaki:2016qgc,Nakamura:2014fxa,Takiwaki:2021dve}.

The  neutrino and gravitational wave signals from three-dimensional SN simulations exhibit clear detectable signatures of hydrodynamic instabilities~\cite{Tamborra:2013laa,Tamborra:2014aua,Tamborra:2014hga,Walk:2018gaw,Walk:2019ier,Walk:2019miz,Lund:2010kh,Lund:2012vm,Kuroda:2017trn,Muller:2014rpb,Takiwaki:2021dve,Shibagaki:2020ksk,Takiwaki:2017tpe,Kuroda:2017trn,Kuroda:2016bjd,Kuroda:2013rga,Andresen:2018aom,Andresen:2016pdt,Pajkos:2020nti,Ott:2012kr,Muller:2019upo,OConnor:2018a,Radice:2019,Mezzacappa:2020,Mezzacappa:2022xmf},  especially in the presence of  the standing accretion shock instability (SASI)~\cite{Blondin:2002sm,Scheck:2007gw}. 
 For example,  the Fourier power spectrum of the neutrino rate expected from a SN burst occurring in our Galaxy is expected to show a sharp peak corresponding to the SASI frequency~\cite{Tamborra:2013laa,Lund:2012vm}, since the neutrino signal carries characteristic time modulations replicating the sloshing of the stalled shock wave during a SASI phase; a similar periodic modulation is also observable in gravitational waves~\cite{Andresen:2016pdt,Shibagaki:2020ksk,Mezzacappa:2022xmf}. 

In the presence of rotation, the physics linked to SASI is currently uncertain.  The detectable modulations of the neutrino signal can be smeared out by rotation, and modulations in the  neutrino spectrogram at frequencies larger than the SASI one appear~\cite{Walk:2018gaw,Walk:2019ier}. A neutrino light-house effect has also been found in the presence of the low $\mathrm{T}/|\mathrm{W}|$ instability developing for very large rotation rates~\cite{Takiwaki:2017tpe,Shibagaki:2020ksk}. The dynamics of the low $\mathrm{T}/|\mathrm{W}|$ instability in the accreting proto-neutron star may be triggered by the buoyancy driven dynamics at mid-latitudes above and below the equatorial plane~\cite{Takiwaki:2021dve}.

These recent developments and the detectable multi-messenger implications call for a deeper analysis of the development of SASI and its interplay with rotation in the equatorial plane.
The efficiency of the advective-acoustic cycle has been investigated in a simple, non-rotating spherical model by means of the linear stability analysis \cite{Foglizzo:2006fu}. It was found that the lowest ($l=1$) mode is the most unstable for a physical range of shock to proto-neutron star radius ratios. 
The axisymmetric component $m=0$ of this mode corresponds to a dipolar side-to-side sloshing motion of the post-shock region, whereas the components $m=\pm1$ correspond to spiral patterns. All $m$-components share the same radial structure and eigenfrequency \citep{Foglizzo:2006fu}, so that the perturbative evolution of a sloshing mode can be described as a linear combination of spiral patterns with different phases and vice-versa \citep{Fernandez:2010}. This decomposition is not possible in a rotating flow because the radial structure and eigenfrequency of the modes depend on their azimuthal wavenumber $m$ measured with respect to the rotation axis, with prograde spiral modes dominating the dynamical evolution \citep{Blondin:2007}. 
Incorporating rotation in a perturbative analysis in 3D
is challenging  because flow gradients along the poloidal direction significantly complicate the set of hydrodynamic equations. By focusing on the 2D dynamics in the equatorial plane and enforcing certain symmetries, angular momentum can be taken into account in a simplified manner, e.g.~a cylindrical model with uniform angular momentum and invariance along the $z$-axis was investigated through a perturbative analysis~\cite{Yamasaki:2007dc} and 2D hydrodynamic simulations~\cite{Kazeroni:2017fup,Blondin:2017}. These studies confirmed the destabilizing impact of rotation on the prograde spiral mode $m=1$ of SASI pointed out in Ref.~\cite{Blondin:2007} and revealed the dominance of the $m=2$ spiral mode for larger rotation rates.

Although the cylindrical setup allows for the self-consistent incorporation of angular momentum, the assumption of a cylindrical geometry is expected to affect the postshock velocity profile (see, e.g.,~Fig.~12 of Ref.~\cite{Kazeroni:2015qca}) and thus the advection time from the shock. This timescale is an important ingredient of the advective-acoustic mechanism driving SASI~\cite{Foglizzo:2006fu,Fernandez:2009,Guilet:2012}. 
The velocity profile of the spherical geometry is preserved in the spherical equatorial approximation used in the 2D simulations presented in  Ref.~\cite{Blondin:2006fx}, which also neglects the flow gradients in the poloidal direction. The consistency of the results obtained in 2D cylindrical, 2D spherical equatorial and 3D spherical geometries has been verified in Ref.~\cite{Blondin:2017}. The difference between the cylindrical and spherical geometries is indeed modest when the shock distance is a few times the radius of the proto-neutron star. 
These results have also been confirmed by 3D numerical simulations of stationary accretion incorporating a more realistic equation of state and neutrino heating~\cite{Iwakami:2009a,Iwakami:2009b}. Such perturbative and numerical studies of toy models involving SASI and rotation have provided insight on the results of hydrodynamical simulations of stellar core collapse in 3D~\cite{Summa:2017wxq, Takiwaki:2021dve}. 

In this work, we take advantage of the different advection times implied by the cylindrical and spherical geometries in the asymptotic limit of  large shock radius to test our understanding of the mechanism of SASI in the presence of rotation, using a perturbative analysis. 
In Sec.~\ref{sec:models}, our two toy models employing spherical equatorial and cylindrical geometries are introduced and the spherical equatorial model is compared to the spherical model without rotation. 
The stationary flow profiles in spherical equatorial and cylindrical geometries are used to infer the expected properties of the advective-acoustic cycles in Sec.~\ref{sec:SF}. The growth rates and oscillation frequencies of the fundamental $m=0$ and $m=1$  modes for the spherical and cylindrical models without rotation are computed numerically and their differences are interpreted physically in Sec.~\ref{sec:eigenmodes}. The impact of rotation on the development of SASI and the relative insensitivity to the geometry of the flow is discussed in Sec.~\ref{sec:rotation}. Concluding remarks are reported in Sec.~\ref{sec:conclusions}.

\section{Cylindrical and spherical equatorial toy models} \label{sec:models}

Given the challenges intrinsic to full-scale hydrodynamic simulations, throughout this work we rely on two toy models of a perfect gas with a uniform specific angular momentum falling in a Newtonian potential set by the proto-neutron star, employing different geometries. 
Firstly, we consider the cylindrical system introduced in Ref.~\cite{Yamasaki:2007dc}. This model is described by the coordinate system $(r, \phi, z)$ where $\phi$ is the azimuthal angle, such that both the accretor and shock front have  cylindrical geometry. We impose invariance along the rotation axis ($z$) such that for the steady state flow, all gradients along this axis are zero ($\partial/\partial z = 0 $) and accretion is restricted to the radial direction ($v_z = 0$). 
We only consider the case where the vertical wave number of the perturbed shock front in the cylindrical system is zero.

Secondly, we investigate a spherical equatorial model similar to that of Refs.~\cite{Blondin:2006fx,Blondin:2017,Abdikamalov:2021} in spherical coordinates $(r, \theta, \phi)$. For the sake of  simplicity, this model assumes  invariance along the axis perpendicular to the equatorial plane ($\partial v_\theta/\partial \theta=0$ for $\theta = \pi/2$), mimicking the cylindrical model in the equatorial plane. Thus, the spherical coordinate system effectively reduces to $(r, \phi)$.  
 We apply the same assumptions of Refs.~\cite{Foglizzo:2006fu,Yamasaki:2007dc} for each model:
\begin{enumerate}
\item The supersonic matter above the shock front is cold and free falling, the pre-shock Mach number is $\mathcal{M}_1\gg 1$;

\item The gravitational potential $\Phi_0 \equiv - GM/r$ is set by the central neutron star with mass $M$, and self-gravity is neglected;

\item The flow is inviscid, and the specific angular momentum ($L$) is uniform and conserved such that $v_\phi = L/r$;

\item The flow is described by a compressible, perfect gas with uniform adiabatic index, $\gamma = 4/3$;

\item The gas is ideal such that $P\propto \rho T_{\rm m}$, where $T_{\rm m}$ is the matter temperature and $\rho$ the density;

\item The effects of photodissociation are neglected across the strong shock. The resulting postshock Mach number is $\mathcal{M}_\mathrm{sh}^2 = (\gamma-1)/{(2\gamma)}$;

\item Neutrino heating is neglected and neutrino cooling is approximated with a cooling function modelled as $\mathcal{L} \propto \rho^{\beta - \alpha} P^\alpha$, where $\alpha = 3/2$ and $\beta=5/2$, following Refs.~\cite{Blondin:2002sm, Foglizzo:2006fu, Yamasaki:2007dc, Blondin:2017}.
\end{enumerate}
The subscript ``1'' corresponds to quantities just above the shock front, while the subscript ``2'' indicates flow quantities just below the shock front. 

The set of equations governing the dynamics is composed by the continuity equation, the entropy equation, and the Euler equation: 
\begin{eqnarray}
\frac{\partial \rho}{\partial t} + \nabla \cdot \rho \vec{v} &=& 0\ , \label{eq:mass_cont}\\
\frac{\partial S}{\partial t} + \vec{v} \cdot \nabla S &=& \frac{\mathcal{L}}{P}\ , \label{eq:entropy}\\
\frac{\partial \vec{v}}{\partial t} + \vec{w} \times \vec{v} + \nabla B &=& \frac{c^2}{\gamma} \nabla S\ , \label{eq:euler}
\end{eqnarray}
where $P$ is the pressure, $c$ the sound speed, $\vec{w} \equiv \nabla \times \vec{v}$  the vorticity, $S$ the dimensionless entropy, and $B$ is the Bernoulli function defined by
\begin{eqnarray}
B(r)&\equiv& 
{\vec{v}\,^2\over2} + {c^2\over\gamma - 1} + \Phi_0\ ,\\
&=&
{v_r^2\over 2}+{L^2\over 2r^2}+{c^2\over\gamma-1}-{GM\over r}\ ,\label{defBernoulli}\\
{\rho\over\rho_{\rm sh}}&=&\left({c\over c_{\rm sh}}\right)^{2\over\gamma-1}{\rm e}^{-(S-S_{\rm sh})}\ .\label{densityprofile}
\end{eqnarray}

The perturbative equations have already been established in cylindrical geometry in Ref.~\cite{Yamasaki:2007dc}, in spherical geometry without rotation in Refs.~\cite{Foglizzo:2006,Foglizzo:2006fu}, and in spherical equatorial geometry without cooling in Ref.~\cite{Abdikamalov:2021}. 

In Appendix~\ref{sec:A4}, we present a unified view of the set of equations parametrized by a geometrical factor $g$ accounting for the cylindrical ($g=1$) or spherical equatorial ($g=2$) character of the flow. In the set of stationary equations, this geometrical factor  mainly impacts the radial distribution of the mass flux:
\begin{eqnarray}
     \rho v_{r} &=& \left({r_{\rm sh}\over r}\right)^g\rho_{\rm sh} v_{r_{\rm sh}}\ , \label{eq:mass_cons}\\ 
    \frac{\partial S}{\partial r} &=& \frac{\mathcal{L}}{P v_r}\ , \label{eq:entropy_cons}\\
    \frac{\partial B}{\partial r}  &=& \frac{\mathcal{L}}{\rho v_r}\ .\label{eq:euler_cons}
\end{eqnarray}

The perturbations of the stationary flow are expressed using the vector $\delta X(r)\equiv(r\delta v_\phi/im,\delta h,\delta S,\delta K/m^2)$ where the perturbation of the mass flux $\delta h$ and the baroclinic combination $\delta K$ of vorticity $\delta w$ and entropy $\delta S$, introduced in Ref.~\cite{Foglizzo:2001}, are defined as
\begin{eqnarray}
\delta h&\equiv&{\delta v_r\over v_r}+{\delta \rho\over\rho}\ ,\\
\delta K&\equiv&-imrv_r\delta w+m^2{c^2\over\gamma}\delta S\ ,
\end{eqnarray}
where $m$ is the azimuthal wavenumber.
The differential system of equations and boundary conditions at the shock ($r_{\rm sh}$) and at the surface of the proto-neutron star ($r_\ast$) are expressed in Appendix~\ref{sec:A4} in the following compact form using the matrix ${\bf M_r}$ and boundary condition vectors $Y^{\rm sh}$ and $Y^{\ast}$:
\begin{eqnarray}
{\partial \delta X\over\partial r}
&=&{\bf M_r}\delta X\ \label{matrix_form}\ ,\\
\delta X_{\rm sh}&=&
Y^{\rm sh}\Delta\zeta\ ,\label{boundarysh}\\
\delta X_{\ast}\cdot Y^{\ast}&=&0\ ,\label{boundaryNS}
\end{eqnarray}
where $\Delta\zeta$ is the shock displacement. The vectors $Y^{\rm sh}$ and $Y^{\ast}$ are introduced in Appendix~\ref{sec:A4} in Eqs.~(\ref{Ysh1})--(\ref{Ysh4}) and Eqs.~(\ref{Yns1})--(\ref{Yns4}), and the matrix ${\bf M_r}$ is defined by Eqs.~(\ref{Mr11})--(\ref{Mr44}).

{\bf Impact of the flow geometry and rotation parameters.}---The  choice of the variable $\delta X$ allows to formulate the boundary value problem (Eqs.~\ref{matrix_form}--\ref{boundaryNS}), where the frequency $\omega$ appears only through the Doppler shifted frequency 
\begin{eqnarray}
\omega^\prime\equiv\omega-mL/r^2\ ,\label{defomegaprime}
\end{eqnarray}
both in the differential system of equations and in the boundary conditions. The geometrical parameter $g$ and the quadratic effect of the centrifugal force $L^2$ appear explicitly only in the third component of $Y^{{\rm sh}}(\omega^\prime,L^2,g)$ (Eq.~\ref{Ysh3}), defining the entropy perturbation at the shock (Eqs.~\ref{eq:BCS0}--\ref{eq:BCS}), as a consequence of the change of slope of the momentum flux across the shock (Eq.~\ref{eq:stat2}). It is remarkable that both $g$ and $L^2$ are explicitly absent from the coefficients of the matrix ${\bf M_r}(\omega^\prime)$
(Eqs.~\ref{Mr11}--\ref{Mr44}) and the inner 
boundary condition $Y^{\ast}(\omega^\prime)$ (Eqs.~\ref{Yns1}--\ref{Yns4}), although they are implicitly present through the equations defining the stationary flow. The main impact of the flow geometry is indeed the advection velocity, analyzed in Sec.~\ref{sec:SF}. The effect of modest rotation through the Doppler shifted frequency $\omega^\prime$ is investigated in Sec.~\ref{sec:rotation}.

\begin{figure}[t!]
\centering
\includegraphics[width=\columnwidth]{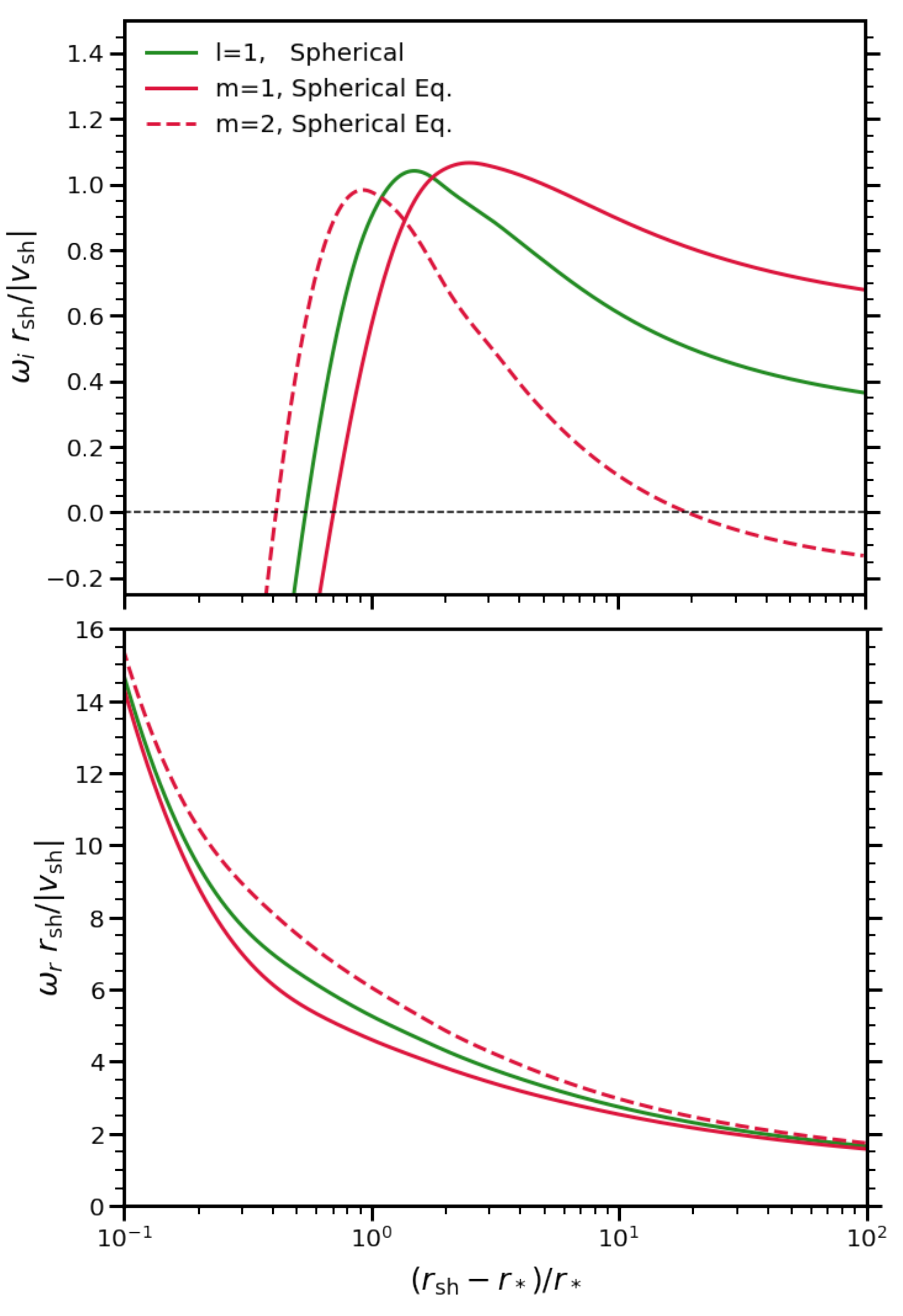}
\caption{The eigenfrequency of the mode $l=1$  in spherical geometry ($m=-1,0,1$, green curve) is intermediate between the eigenfrequencies of the  modes $m=1$ and $m=2$ in  spherical equatorial geometry (solid and dashed red curves, respectively) in the absence of rotation. The growth rate of the  mode $m=1$ is the largest for $r_{\rm sh}/r_\ast \gtrsim 2.37$, while the  mode $m=2$ is dominant for smaller ratios. The oscillation frequencies are remarkably similar.}
\label{fig:sph_comparison}
\end{figure}

{\bf Quantitative impact of neglecting the poloidal structure of the perturbed flow.}---Before exploiting the geometrical difference between the cylindrical and spherical equatorial models, with and without rotation, we evaluate the quantitative impact of neglecting the poloidal structure of the perturbations. As in Ref.~\cite{Yamasaki:2007dc}, we note that the eigenvalue $-l(l+1)$ of the Laplacian operator in spherical geometry is replaced by $-m^2$ in the spherical equatorial and cylindrical geometries considered here. 

Without rotation, the {\it only} difference between the differential system in Eqs.~(\ref{matrix_form})--(\ref{boundaryNS}), defining the eigenfrequencies in spherical equatorial geometry, and the differential system  in spherical geometry of  Ref.~\cite{Foglizzo:2006fu} is in the coefficient $M_r^{21}$ (see Eq.~\ref{Mr21}) expressing the cylindrical Lamb frequency $\omega_{\rm Lamb}$ as
\begin{eqnarray}
\omega_{\rm Lamb}^2\equiv {m^2\over r^2}(c^2-v_r^2) ,\label{defLamb}
\end{eqnarray}
instead of the spherical formula $\omega_{\rm Lamb}^2\equiv l(l+1)(c^2-v_r^2)/r^2$ (Eq.~14 in Ref.~\cite{Foglizzo:2006fu}).
The impact of this difference is measured without rotation in Fig.~\ref{fig:sph_comparison} by comparing the $l=1$ eigenfrequencies in the spherical model and the $m=1$ frequencies in the spherical equatorial model.
We explore a wide range of length scales of the post-shock layer ($ 1 \leq (r_{\mathrm{sh}} - r_\ast) / r_\ast \leq 100$) by tuning the strength of the neutrino cooling function $\mathcal{L}$ to obtain the desired $r_\ast$ to $r_\mathrm{sh}$ ratio. The eigenmodes are then found with a shooting method using a steepest descent algorithm.

As expected the growth rate and oscillation frequency of the mode $l=1$ in spherical geometry is independent of $m=-1,0,1$ \citep{Foglizzo:2006fu} and intermediate between the $m=1,2$ modes in equatorial-spherical geometry, since it corresponds to the same mathematical system (Eqs.~\ref{matrix_form}--\ref{boundaryNS}) with $m=\sqrt{2}$. The difference of growth rates between the spherical and spherical equatorial geometries may have been too small to be noticed in the numerical simulations presented in Ref.~\cite{Blondin:2017} with $r_{\rm sh}/r_\ast=5$ and very modest rotation (see their Fig.~9). The difference between the growth rates is most pronounced in the asymptotic regime of  large shock radius, see
 the curves corresponding to  $m=1$ and $l=1$ in Fig.~\ref{fig:sph_comparison}, while the oscillation frequencies are asymptotically identical. The comparison between the curves corresponding to $m=1$ and $m=2$ in Fig.~\ref{fig:sph_comparison} indicates that the spherical equatorial approximation preserves the 
property that smaller angular scales are favored for smaller distances between the shock and the proto-neutron star surface,
like in spherical geometry~\citep{Foglizzo:2006fu}.

The  identification of the separate roles of $m^2$ in the Lamb frequency and $m$ in the Doppler shifted frequency suggests that a better approximation of the eigenfrequencies in spherical geometry with rotation could be obtained in future work. This could be implemented by introducing an additional parameter $l$ for the angular degree of the mode in the same mathematical system (Eqs.~\ref{matrix_form}--\ref{boundaryNS}), replacing $m^2$ by $l(l+1)$ in Eq.~(\ref{Mr21}) and keeping $m$ as the azimuthal wavenumber everywhere else.

\section{Impact of  flow geometry on  advection timescale and  entropic-acoustic coupling} \label{sec:SF}

\begin{figure}[b]
\centering
\includegraphics[width=\columnwidth]{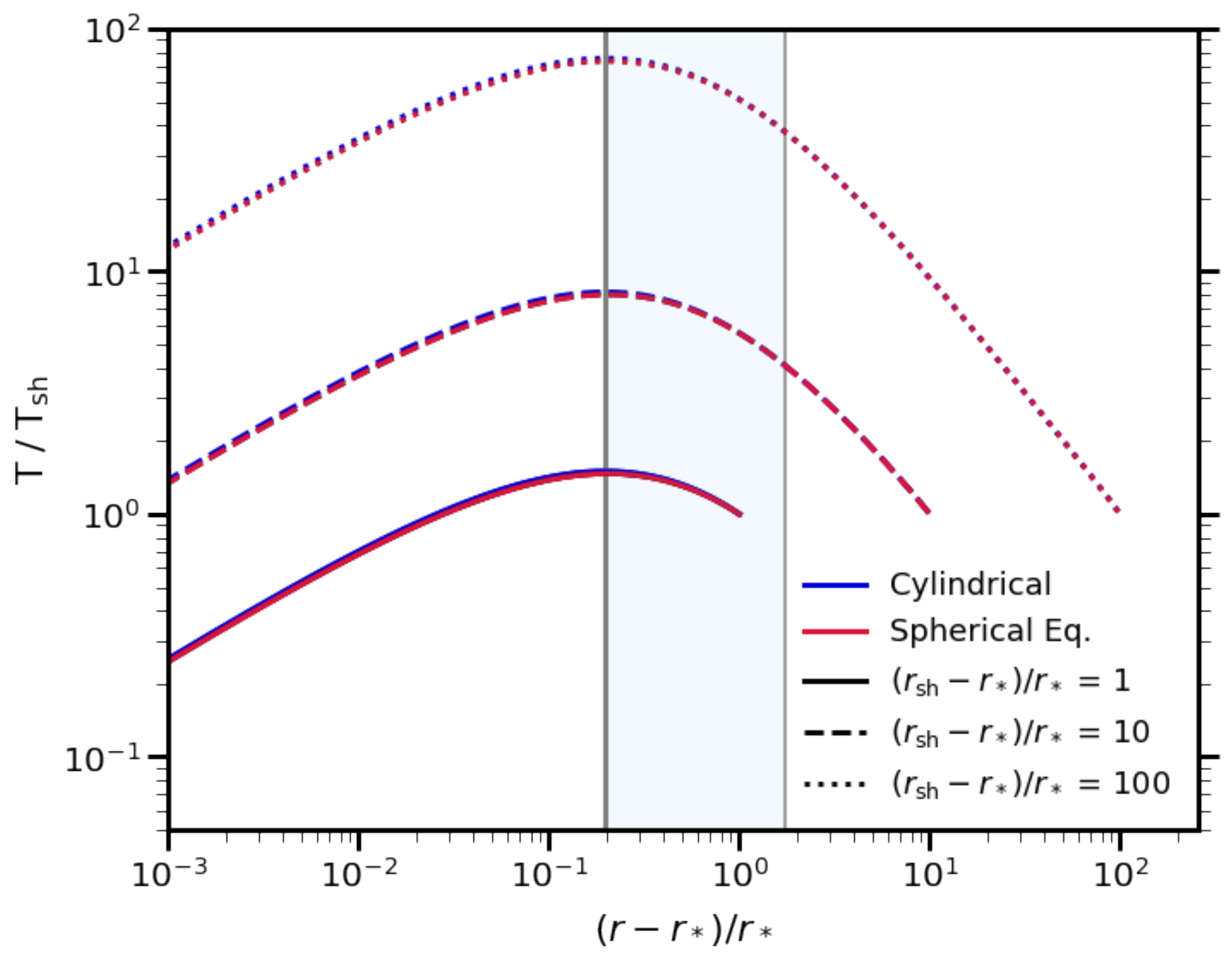}
\caption{Radial profile of the matter temperature of the stationary flow  for three ratios of the shock to proto-neutron star radius without rotation for the cylindrical (in blue) and spherical equatorial (in red) models. The temperature maximum marked by the thick vertical gray line  roughly lies at the same radius $r_{\rm peak}\sim 1.2\, r_\ast$ in both geometries. The width of the temperature peak (indicated by the light blue region), $\Delta r_{\rm peak}\sim \, 1.5 r_\ast$ and bound by the thin gray vertical line at $r_{T/2}$ and the temperature peak, is also independent of the flow geometry. Note that the curves are almost identical for both geometries for our  model parameters.}
\label{fig:Temp_Profile}
\end{figure}

Figure~\ref{fig:Temp_Profile} represents the temperature profiles in spherical and cylindrical geometries, without rotation, for three values of the ratio of the shock to the neutron star radius. 
It shows that in the asymptotic limit of a large shock radius, the subsonic flow below the stationary shock is approximately adiabatic from the shock down to the cooling layer, where neutrino emission becomes dominant. Equations~(\ref{eq:entropy_cons})--(\ref{eq:euler_cons}) with ${\cal L}=0$ suggest that the adiabatic part of the flow is characterized by a uniform entropy $S=S_{\rm sh}$ and  a uniform Bernoulli constant $B_{\rm sh}\equiv B(r_{\rm sh})$ such that: 
\begin{eqnarray}
B_{\rm sh}
&=&{c^2\over \gamma-1}\left( 1+{\gamma-1\over 2}{\cal M}^2\right) - {GM\over r}+{L^2\over 2r^2}\ .\label{soundprofile}
\end{eqnarray}
Equations~(\ref{densityprofile}) and (\ref{soundprofile}) indicate that the profiles of sound speed and density in the subsonic adiabatic part of the flow are mainly ruled by the balance between gravitational and pressure forces, $\partial c^2/\partial r\sim -(\gamma-1)GM/r^2$, with a minor contribution from the ram pressure and a quadratic effect of the centrifugal force. 
The radial velocity is deduced from the conservation of the mass flux and entropy and depends on the geometry through the parameter $g$ as follows:
\begin{eqnarray}
 {v_r\over v_{\rm sh}}  &=&
\left({c\over c_{\rm sh}}\right)^{-{2\over \gamma-1}} \left({r_{\rm sh}\over r}\right)^g\ \\
&\sim & \left({r\over r_{\rm sh}}\right)^{{1\over\gamma-1}-g}\ ,\label{velocityprofile}\\
{{\cal M}^2\over{\cal M}_{\rm sh}^2}&\sim & \left({r\over r_{\rm sh}}\right)^{{\gamma+1\over\gamma-1}-2g}\ .\label{profM2}
\end{eqnarray}
Equation~(\ref{velocityprofile}) implies higher radial fluid velocities in the spherical model compared to the cylindrical one, due to a stronger nozzle effect in the subsonic regime. 
For $\gamma=4/3$, the correction associated to the subsonic Mach number in Eq.~(\ref{soundprofile}) involves a power $(7-2g)$ of the radius according to Eq.~(\ref{profM2}), equal to $3$ in spherical geometry and $5$ in cylindrical geometry.
Cooling processes become dominant in the vicinity of the neutron star, resulting in the decrease of the Bernoulli parameter and  entropy. In the stationary toy model, where the radial velocity vanishes at the surface of the neutron star while keeping a constant mass flux, the diverging density with a constant pressure implies a decrease of the sound speed $c^2=\gamma P/\rho$ to zero at the neutron star surface, after reaching a maximum at an intermediate radius $r_{\rm peak}$, as shown in Fig.~\ref{fig:Temp_Profile}. 
The fact that $r_{\rm peak}\sim 1.2 r_\ast$ seems independent of the shock distance and the geometry is not obvious a priori. 
The width of the temperature peak, 
$\Delta r_{\rm peak}\equiv r_{T/2} - r_{\rm peak}$, 
in the adiabatic part of the flow is defined as the region between the temperature maximum and the point $r_{T/2}$ where the temperature  decreases to half its maximum value, as shown in Fig.~\ref{fig:Temp_Profile}. 
 We note that
$\Delta r_{\rm peak}\sim 1.5 r_\ast$ regardless of the value of $r_{\rm sh}$ or the choice of geometry.

\begin{figure}[t]
\centering
\includegraphics[width= \columnwidth]{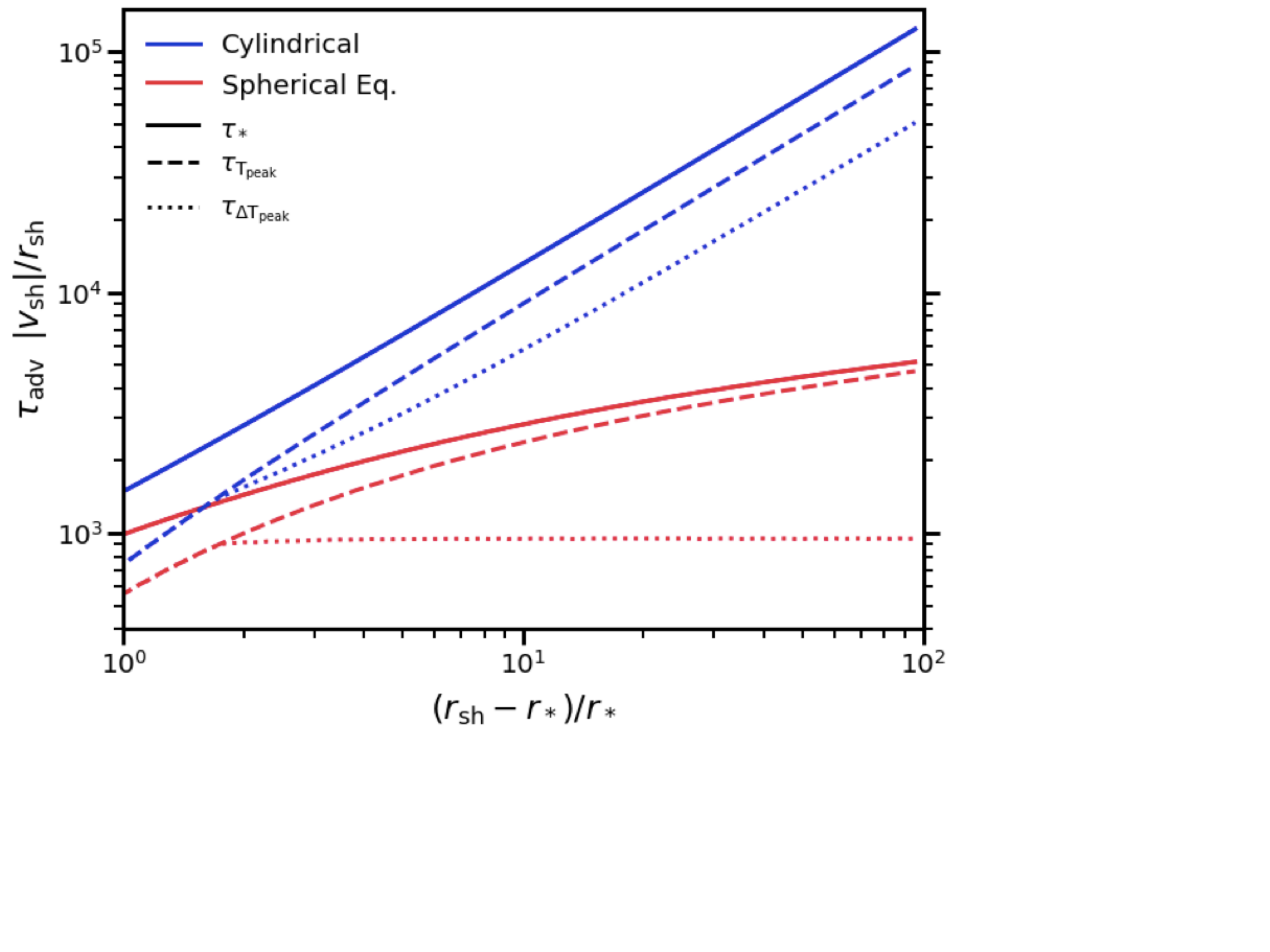}
\caption{Radial profile of the advection time of the fluid from the shock to the inner boundary ($\tau_\ast$), from the shock to the the temperature peak ($\tau_{T_{\rm peak}}$), and through the width of the temperature peak ($\tau_{\Delta T_{\rm peak}}$) for the cylindrical and spherical equatorial models (in blue and red, respectively) in the absence of rotation. The advection timescales reflect the velocity profile of each model ($v_r \propto r$ in the spherical equatorial  model and $v_r \propto r^2$ in the cylindrical model). In the cylindrical model, $\tau_{\Delta T_{\rm peak}}$ is almost half of the total advection time, whereas in the spherical equatorial  model, it is only a small fraction. As the width of the temperature peak is constant between the models and with increasing $r_\mathrm{sh}$ (see Fig.~\ref{fig:Temp_Profile}), this implies that a strong entropic-acoustic coupling is favored in the spherical equatorial  model. 
}
\label{fig:Tadv}
\end{figure}

Although the temperature profile in  Fig.~\ref{fig:Temp_Profile} shows no significant difference between the two models, the geometrical dependence of the velocity profile impacts the efficiency of the entropic-acoustic coupling in the region of adiabatic compression. As advected perturbations of entropy travel across the width of the temperature peak, the entropic-acoustic coupling can contribute to generate coherent SASI oscillations~\citep{Foglizzo:2000, Foglizzo:2009}. Assuming that this is the dominant contribution, the fundamental frequency  of SASI oscillations is approximately set by the advection time $\tau_{\rm peak}$ from the perturbed shock down to $r_{\rm peak}$ ($\omega_{\rm SASI}\sim 2\pi/\tau_{\rm peak}$). We note, however, that this relation is approximate since it neglects the acoustic time from $r_{\rm peak}$ to $r_{\rm sh}$, as well as the possibility of a significant phase shift or  contribution of the purely acoustic cycle. Figures~2, 8 and 9 of Ref.~\cite{Foglizzo:2009} illustrate these contributions within a simplified plane parallel setup. The timescale $\tau_{\rm peak}$ is shown in the cylindrical and spherical geometries in Fig.~\ref{fig:Tadv}. 
The advection time $\tau_\nabla$ across the temperature peak, also plotted in Fig.~\ref{fig:Tadv}, is measured from $r_{T/2}$ to $r_{\rm peak}$. It defines a frequency cutoff $\omega_\nabla\equiv1/\tau_\nabla$ above which the coupling efficiency is reduced by phase mixing (Eq.~44 in Ref.~\cite{Foglizzo:2009}). With $v_r\propto r^2$ in the cylindrical model and $v_r\propto r$
in the spherical model, the ratio $\omega/\omega_\nabla$ is estimated as follows:
\begin{eqnarray}
{\omega_{\rm SASI}\over\omega_\nabla}&\equiv&2\pi
{\int_{\rm T/2}^{\rm peak}{{\rm d}r/|v_r|}
\over
\int_{\rm sh}^{\rm peak}{{\rm d}r/|v_r|}}\ \\
&=&
2\pi{\Delta r_{\rm peak}\over r_{T/2}}
{r_{\rm sh}\over r_{\rm sh}-r_{\rm peak}}
\;\;({\rm cylindrical})\ ,\label{tadvcyl}\\
&=&2\pi
{\log{(r_{T/2}/r_{\rm peak}) }
\over
\log{(r_{\rm sh}/r_{\rm peak})}}
\;\;({\rm spherical})\ .\label{tadvsph}
\end{eqnarray}
With $r_{\rm peak}\sim 1.2 r_\ast$ and $\Delta r_{\rm peak}\sim 1.5 r_\ast$:
\begin{eqnarray}
{\omega_{\rm SASI}\over\omega_\nabla}&\sim&
{3.5 
r_{\rm sh}\over r_{\rm sh}-r_\ast}
\;\;({\rm cylindrical})\ ,\\
{\omega_{\rm SASI}\over\omega_\nabla}&\sim&
{5.1
\over
\log{(r_{\rm sh}/r_\ast)}}
\;\;({\rm spherical})\ .
\end{eqnarray}
In the asymptotic limit $r_{\rm sh}\gg r_\ast$, the phase mixing is inevitable in cylindrical geometry ($\omega_{\rm SASI}/\omega_\nabla > 3.5$) and can be modest in spherical geometry ($\omega_{\rm SASI}/\omega_\nabla \sim 1$), although both models have nearly identical temperature profiles (Fig.~\ref{fig:Temp_Profile}).

We note  that our estimates of the advection time in Eqs.~(\ref{tadvcyl})--(\ref{tadvsph}) specifically depend  on the value of the adiabatic index in Eq.~(\ref{velocityprofile}). For reference, an increase of the adiabatic index from $\gamma=4/3$ to $\gamma=3/2$ would lead to a similar decrease of the exponent in Eq.~(\ref{velocityprofile}) as the one  produced by a change of the geometrical factor from $g=1$ to $g=2$. 

As confirmed by Fig.~\ref{fig:Tadv}, the advection time is too short in the cylindrical model to benefit from the adiabatic compression near the neutron star surface.
We thus expect the spherical model to develop stronger SASI oscillation compared to the cylindrical model, solely based on the comparison between the stationary flow profiles. This expectation is confirmed in the next section through the numerical calculation of the eigenfrequencies.


\section{Asymptotic properties of the SASI eigenmodes without rotation}\label{sec:eigenmodes}

\begin{figure}[]
\centering
\includegraphics[width=\columnwidth]{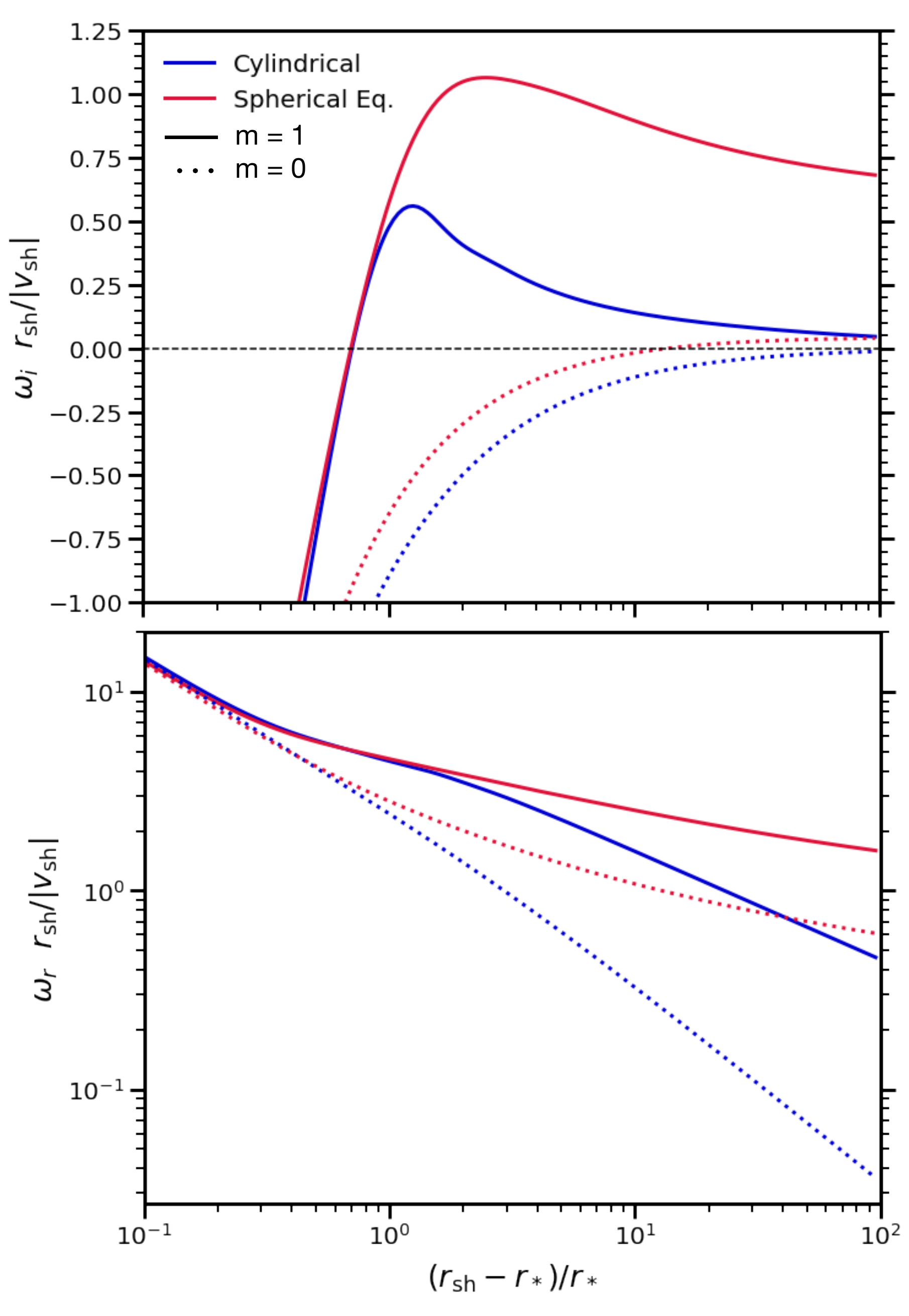}
\caption {Radial profiles of the growth rate (top panel) and oscillation frequency (bottom panel) for the   eigenmodes $m=1$ (solid) and $m=0$ (dotted) for the spherical equatorial  and cylindrical models (in red and blue, respectively) in the absence of rotation. Expressed in units of $|v_{\rm sh}|/r_{\rm sh}$, the growth rate in the spherical equatorial  model remains significant up to large shock radii.
Both the real and imaginary components of the mode $m=1$ decrease steeply in the cylindrical model, as the  shock radius increases. 
The mode $m=0$ becomes slightly unstable in the spherical equatorial  model, while this is never the case in the cylindrical one. This illustrates the strong contribution of entropic-acoustic coupling to the development of the instability in the spherical model.  }
\label{fig:Eigenfrequency}
\end{figure}

The growth rate $\omega_i$ and oscillation frequency $\omega_r$ of the fundamental modes $m=0, 1$ are displayed in Fig.~\ref{fig:Eigenfrequency} for the spherical and cylindrical models, with variations in $r_\mathrm{sh}$. The mode $m=1$ is the most unstable in both models for the entire range of considered length scales. However, the growth rate and oscillation frequency of both models decrease with increased $r_\mathrm{sh}$, when expressed in units of $|v_{\rm sh}|/r_{\rm sh}$.

Figure~\ref{fig:Eigenfrequency} provides evidence for the aforementioned  differences due to the geometry between the two models. Expressed in units of $|v_{\rm sh}|/r_{\rm sh}$, the mode $m=1$ reaches a notably larger growth rate in spherical geometry than in  cylindrical geometry. We note that the same plot expressed in units of the advection time down to the neutron star surface could be misleading, as analyzed in Appendix~\ref{sec:A_Q}.
Regardless of the units we use to measure the eigenfrequency, the mode $m=0$ becomes unstable at large shock radii in the spherical model, while this is not the case for  the cylindrical one. 
Furthermore, the oscillation frequency (in units of $|v_{\rm sh}|/r_{\rm sh}$) decreases in both models as the size of the cavity is increased. However, both modes have a higher oscillation frequency in the spherical model compared to the cylindrical one, the $m=1$ ($m=0$) mode reaching a value up to $3.5$ ($10.7$) times larger for 
$(r_{\rm{sh}} - r_\ast) / r_\ast = 10^2$. 
These properties can be explained by a significant contribution of the entropic-acoustic coupling to the development of the instability of the modes $m=0$ and $m=1$ in the spherical model.

\begin{figure}[]
\centering
\includegraphics[width=\columnwidth]{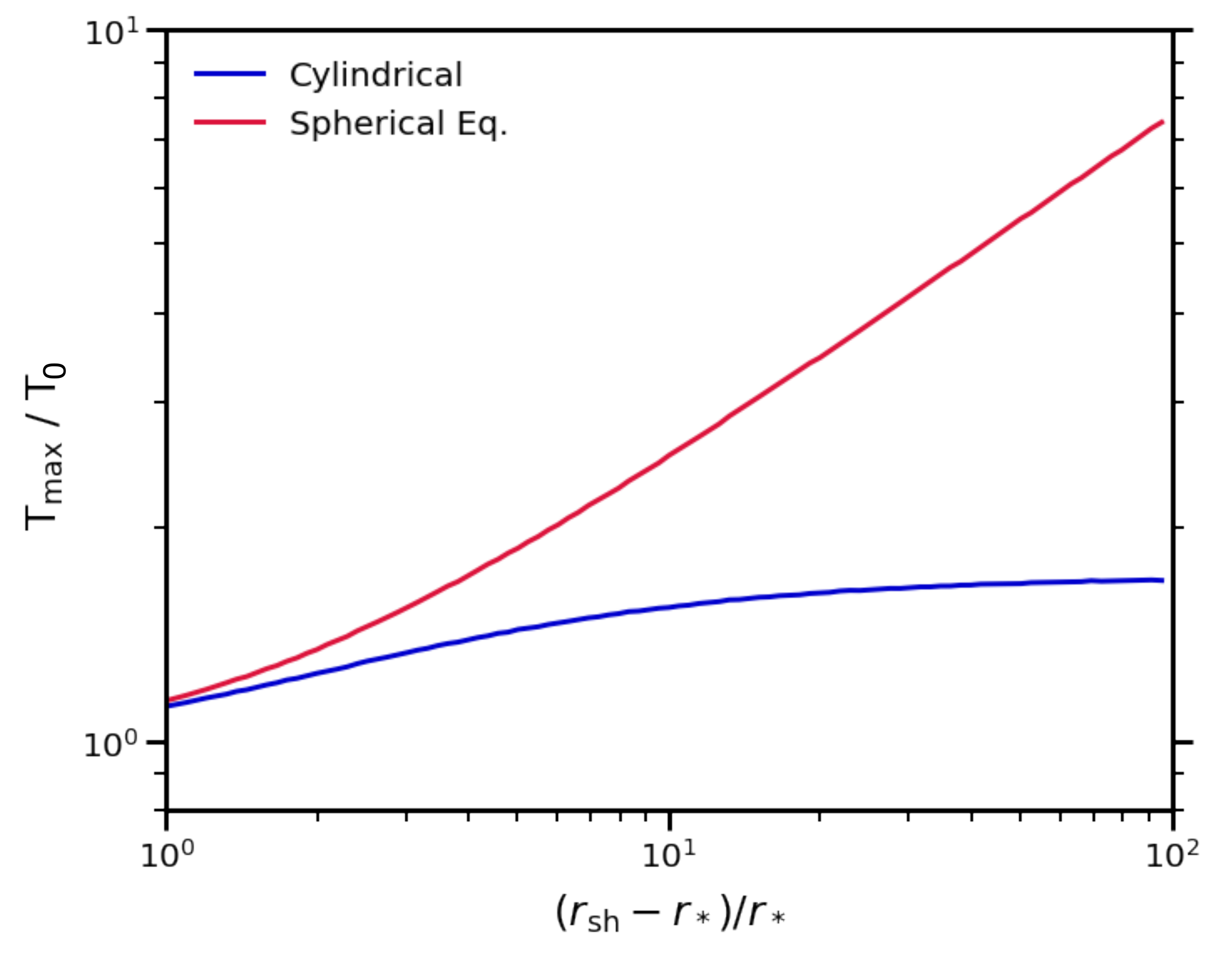}
\caption{Ratio between the maximum temperature $T_{\rm peak}$ and the temperature $T_0$ at the radius $r_0$ reached by an entropy wave advected from the shock during half a period of oscillation $\pi/\omega_r$, for the fundamental eigenmode $m=1$ for the cylindrical and spherical equatorial  models (in blue and red, respectively). This ratio gives the maximum temperature change experienced by an entropy perturbation advected across the region of adiabatic compression. For increasing shock radii, this ratio increases in the spherical equatorial  model, while it is asymptotically bounded in the cylindrical one. }
\label{fig:T_ratio}
\end{figure}

We further test the physical scenario introduced in Sec.~\ref{sec:SF} by measuring  the ratio between the temperature peak $T_{\rm peak}$ and the temperature $T_0$ at the radius $r_0$ reached by an entropy wave advected from the shock during half a period of oscillation $\pi/\omega_r$, as shown  in Fig.~\ref{fig:T_ratio}.
This ratio gives the maximum temperature change along an entropy perturbation with constant sign while advected from the shock to the peak of adiabatic compression. In the cylindrical geometry, this ratio remains limited by $\sim 1.8$ for asymptotically large cavities, while it increases continuously to seemingly arbitrarily large values for the spherical model.

The divergence of the ratio $T_{\rm peak}/T_{0}\propto (r_{\rm sh}/r_\ast)^{1/2}$ for $m=1$ perturbations in spherical geometry, apparent in Fig.~\ref{fig:T_ratio}, suggests a large amplification factor ($|{\cal Q}|\gg1$) of the entropic-acoustic cycle described in Ref.~\cite{Foglizzo:2000}, scaling like a power law of $T_{\rm peak}/T_{0}$. When the entropic-acoustic amplification $|{\cal Q}|$ is large, the contribution of the purely acoustic cycle can be neglected and the growth rate of the advective-acoustic cycle can be approximated by 
\begin{eqnarray}
\omega_i\sim {1\over\tau_{\rm peak}}{\log}|{\cal Q}|\ ,
\end{eqnarray}
where $\tau_{\rm peak}$ is the advection timescale from $r_{\rm sh}$ down to the temperature peak $r_{\rm peak}$ \citep{Foglizzo:2006fu}. 
We note from Eq.~(\ref{velocityprofile}) and Fig.~\ref{fig:Tadv} that the advection time increases like $(r_{\rm sh}/|v_{\rm sh}|)\log(r_{\rm sh}/r_\ast)$ in the spherical model. Combined with the logarithmic increase of $\log |{\cal Q}|$, this is consistent with the saturation of the spherical growth rate $\omega_i r_{\rm sh}/|v_{\rm sh}|$ in Fig.~\ref{fig:Eigenfrequency}. 

Expressing the eigenfrequencies in units of the advection time down to the neutron star surface would produce misleading conclusions, as explained in Appendix~\ref{sec:A_Q}. The more pronounced decrease of the cylindrical growth rate in Fig.~\ref{fig:Eigenfrequency} is in line with the longer advection time, but a more quantitative prediction is precluded by the fact that when the entropic-acoustic cycle is weak, the contribution of the purely acoustic cycle becomes non negligible \citep{Foglizzo:2000,Foglizzo:2006fu}. In this case, the region of most effective advective-acoustic coupling setting the advection timescale may not be determined by the radius of maximum temperature, but rather by the most constructive interference between the vortical-acoustic coupling and the purely acoustic cycle~\cite{Foglizzo:2009,Fernandez:2009}.


\section{Impact of rotation on SASI eigenmodes}\label{sec:rotation}

\begin{figure}[]
\centering
\includegraphics[width=\columnwidth]{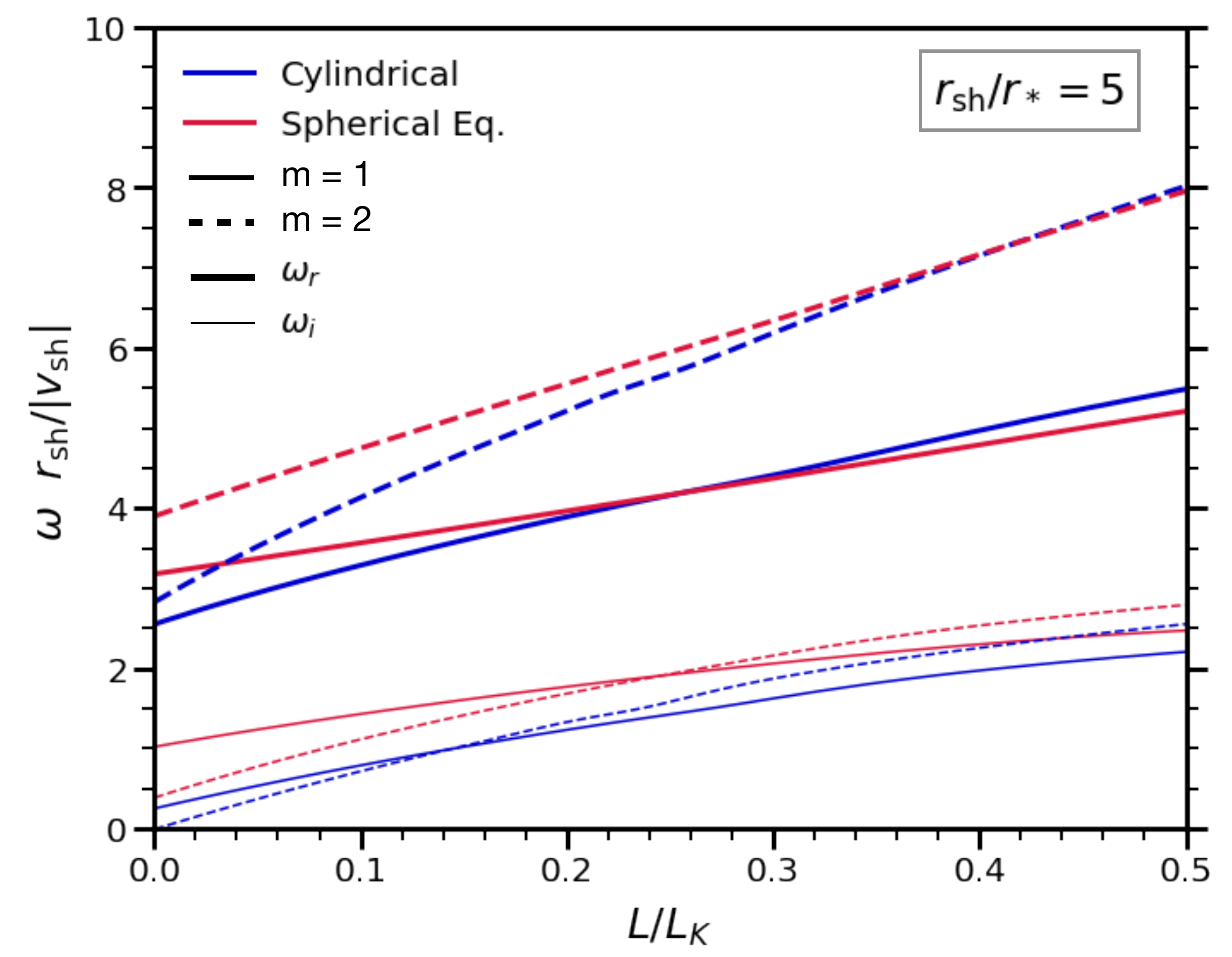}
\caption{Growth rate (thin lines) and oscillation frequency (thick lines) of the modes $m=1$ (solid lines) and $m=2$ (dashed lines) in cylindrical (blue) and spherical equatorial (red) geometries for $r_{\rm sh}/r_{\ast}=5$. The specific angular momentum $L$ is measured in units of the Keplerian limit $L_{\rm K}$ at $r_\ast$. The rotational destabilization of SASI is approximately linear and similar in both geometries.}
\label{fig:Lk_plot}
\end{figure}
Throughout this  section, the effect of rotation on the growth of eigenmodes of our two  toy models is investigated by varying the specific angular momentum ($L$) as well as the shock radius ($r_{\rm sh}$). As the centrifugal force induced by non-zero rotational velocity is much smaller than the gravitational force in our models, the advection time is not  affected by rotation strongly. Thus, the stationary flow quantities introduced  in Sec.~\ref{sec:SF} are only marginally corrected by $L$. 
As noted for cylindrical geometry in Ref.~\cite{Yamasaki:2007dc} and confirmed in spherical geometry by Ref.~\cite{Blondin:2017}, both the growth rate and the oscillation frequency increase approximately linearly with the specific angular momentum. 
This linear increase is confirmed by our perturbative analysis in both geometries, as seen in Fig.~\ref{fig:Lk_plot} for $r_{\rm sh}/r_{\ast}=5$.
This figure also confirms that when the angular momentum exceeds a threshold, the mode $m=2$ becomes more unstable than the mode $m=1$. This threshold is $L/L_{\rm K}\sim 15\%$ in cylindrical geometry and $\sim 25\%$ in spherical equatorial geometry. 
\begin{figure}
\centering
\includegraphics[width=\columnwidth]{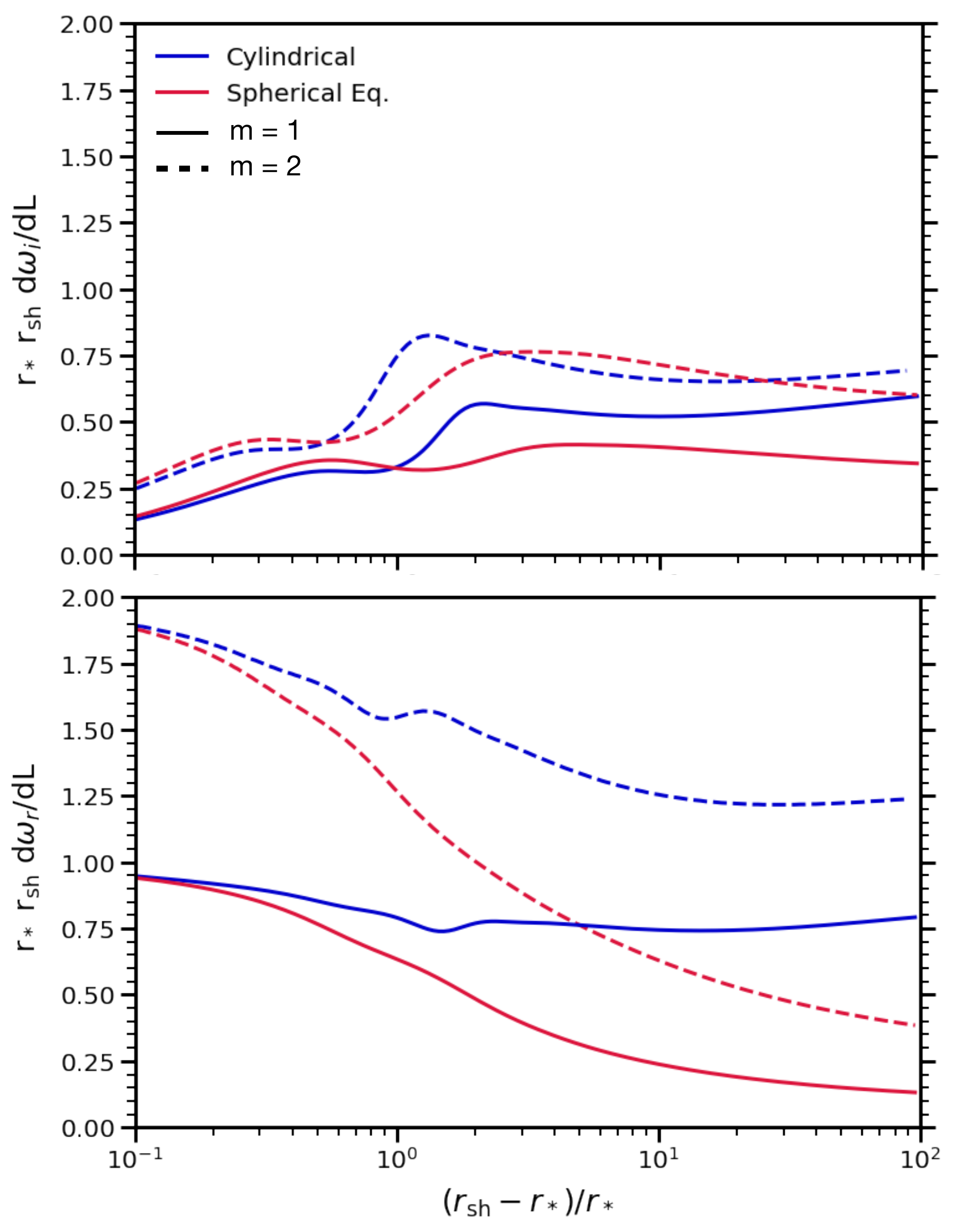}
\caption{Radial profile of the rate of change of the growth rate (top) and oscillation frequency (bottom) of the fundamental modes $m=1$ (solid lines) and $m=2$ (dashed lines)  with increasing rotational velocity, multiplied by the characteristic length scale $r_{\rm sh} r_\ast$. This quantity is relatively constant in both models for $r_{\rm sh} > 3r_\ast$. 
}
\label{fig:Rotation}
\end{figure}
The linear increase of the oscillation frequency was characterized in Ref.~\cite{Blondin:2017} using the rotation frequency at a characteristic propagation radius $r_{\rm p}$ defined by
\begin{equation}
    \frac{d\omega_r}{dL} \sim{1\over r_{\rm p}^2}\ .
\end{equation}
Our Fig.~\ref{fig:Rotation} expands this result by showing the rotational sensitivity of the fundamental eigenmode scaled with the characteristic length $r_\mathrm{sh} r_\ast$ for each geometrical system, as it transitions from a state of zero rotation to one with small angular momentum. 
The dimensionless complex quantity $\mathcal{C}$, defined as
\begin{eqnarray}
 \mathcal{C}
\equiv {1\over r_{\rm sh}  r_\ast}
{{\rm d}\omega\over {\rm d}L}\ ,
\end{eqnarray}
is plotted in Fig.~\ref{fig:Rotation}.
The propagation radius is deduced from the real part $\mathcal{C}_r$ of $\mathcal{C}$ according to
\begin{eqnarray}
    {r_{\rm p}\over r_{\rm sh}}= {1\over \mathcal{C}_r^{1\over 2}} \left({r_\ast\over r_{\rm sh}}\right)^{1\over 2}\ .
\end{eqnarray}
For $r_{\rm sh}/r_\ast=5$, we read $\mathcal{C}\sim(0.35,0.42)$ from Fig.~\ref{fig:Rotation} and recover  $r_{\rm p}/r_{\rm sh}\sim 0.756 $  consistent with the value $0.75$ deduced from the numerical simulations presented in Ref.~\cite{Blondin:2017} (see their Fig.~5). 

We note from Fig.~\ref{fig:Rotation} that $\mathcal{C}$ is relatively constant for large shock radii. For example, we find   $\mathcal{C}\sim (0.15,0.35)$ for $m=1$ in the spherical equatorial model.
The propagation radius $r_{\rm p}\sim 2.6 (r_\ast r_{\rm sh})^{1/2}$ is thus asymptotically proportional to the geometric mean of the shock and neutron star radii.

\begin{figure}[h!]
\centering
\includegraphics[width=\columnwidth]{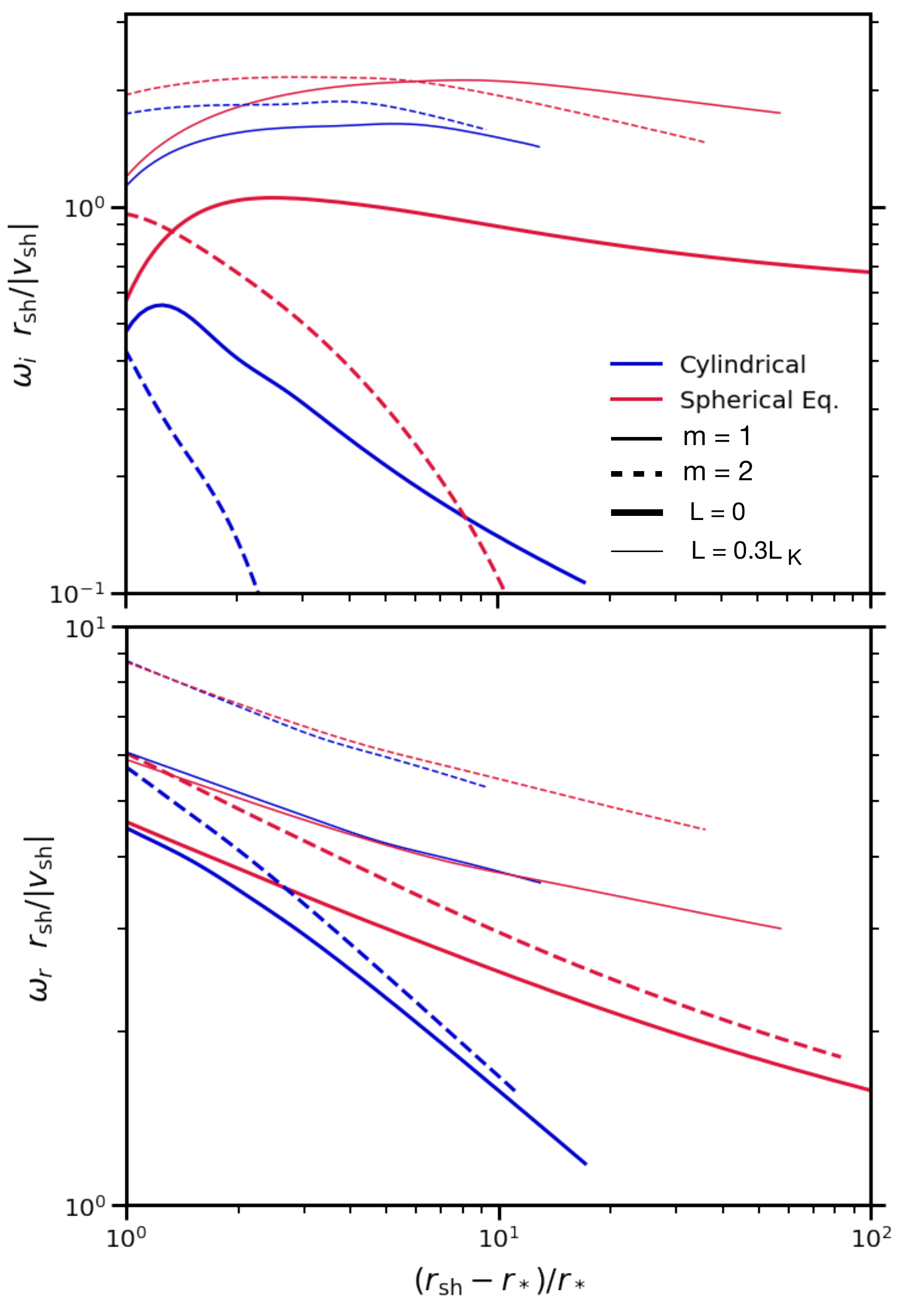}
\caption{Growth rate (top) and oscillation frequency (bottom) of the fundamental modes $m=1$ (solid lines) and $m=2$ (dashed lines) in the cylindrical (blue) and spherical equatorial (red) geometries, for a rotating progenitor with $L=0.3 L_{\rm K}$ (thin lines). The eigenfrequencies of the non-rotating flow are reproduced from Fig.~\ref{fig:Eigenfrequency} for comparison (thick lines).}
\label{fig:OMEGA_PLOT}
\end{figure}
To further investigate the effect of rotation on the development of the eigenmodes in asymptotically large systems, we select a reference angular momentum and compare the growth rate and oscillation frequency of the fundamental mode of the instability to the case of zero rotation. Figure~\ref{fig:OMEGA_PLOT} shows the eigenfrequencies of the modes $m=1$ and $m=2$ for both models as  functions of the cavity size for $L=0.3 L_{\rm K}$, where $L_{\rm K}$ is the Keplerian specific angular momentum defined at $r_\ast$. The eigenfrequencies corresponding to $L=0$ for the mode $m=1$ are reproduced from Fig.~\ref{fig:Eigenfrequency} for comparison. The instability is dominated by the mode $m=1$ when the shock radius exceeds a threshold, which increases with the rotation rate. 

When  the angular momentum increases from $L=0$ to $L=0.3 L_{\rm K}$, Fig.~\ref{fig:OMEGA_PLOT} shows that the geometrical dependence of the oscillation frequency disappears almost completely. An offset remains visible between the growth rates of the two geometries.

\begin{figure}[t!]
\centering
\includegraphics[width=\columnwidth]{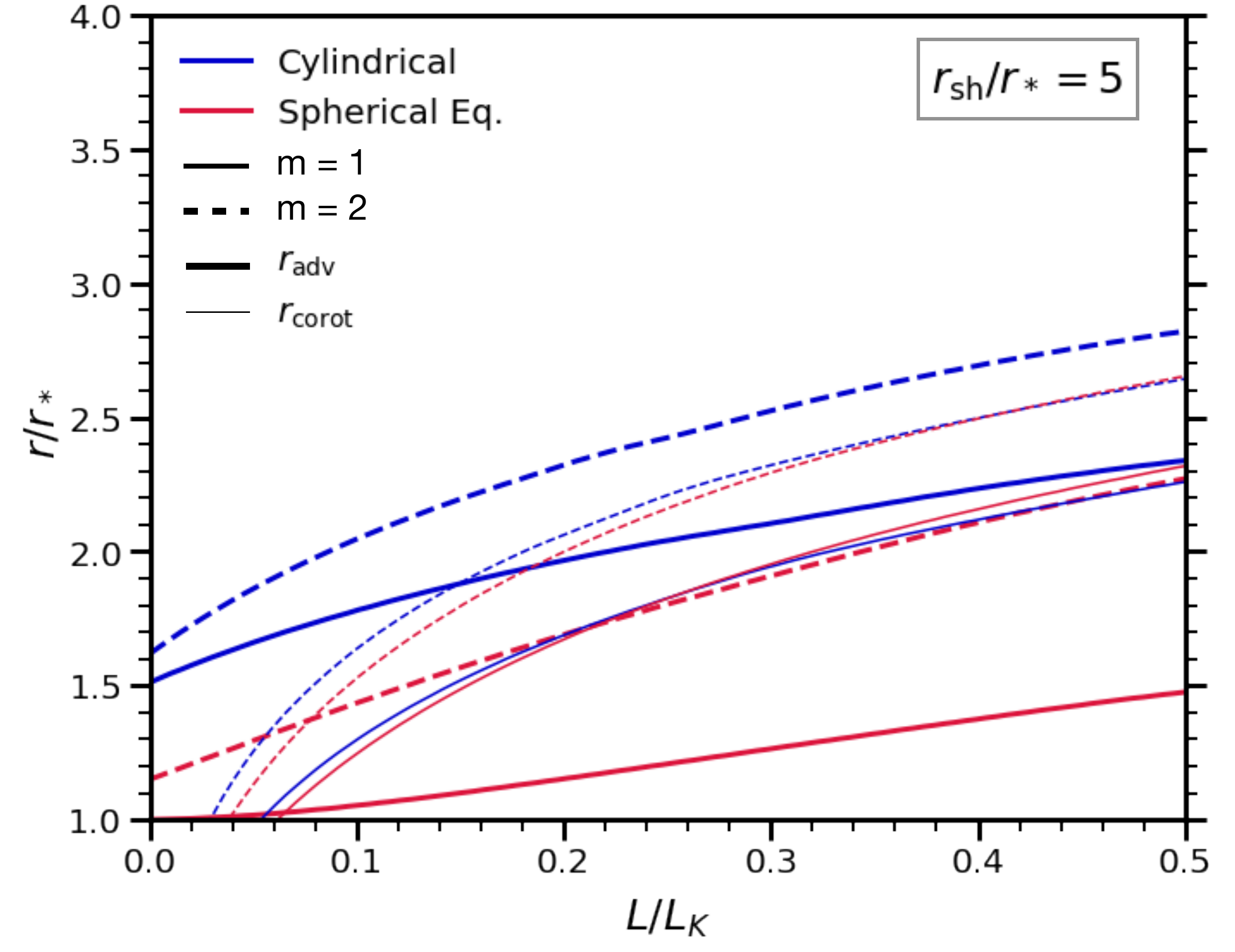}
\caption{Position of the corotation radius $r_{\rm corot}$ (thin lines) and the advection radius $r_{\rm adv}$ (thick lines) reached from the shock after one oscillation period of the fundamental modes $m=1$ and $m=2$ in the cylindrical and spherical equatorial geometries (in blue and red respectively), for $r_{\rm sh}/r_\ast=5$. }
\label{fig:radv}
\end{figure}
With rotation, the weak dependence of the SASI properties on the flow geometry compared to the non-rotating case questions our understanding of the role of advection in the instability mechanism. 
We measure in Fig.~\ref{fig:radv} the radius $r_{\rm adv}$ reached by the gas advected from the shock after one oscillation period $2\pi/\omega_r$ for the modes $m=1$ and $m=2$. This radius is also compared to the corotation radius. The rotational destabilization of SASI could be related to the existence of a corotation radius where the Doppler shifted frequency $\omega^\prime$ vanishes. The low frequency $\omega^\prime$ near corotation could  favor  a large radial wavelength and thus an efficient advective-acoustic coupling, avoiding the damping effect of phase mixing acting in non-rotating flows as described in Ref.~\cite{Foglizzo:2009}. We remark in Fig.~\ref{fig:radv} that the advection radius is always above the corotation radius in cylindrical geometry, whereas it can be significantly below the corotation radius in spherical equatorial geometry. The geometrical effects distinguishing between the cylindrical and spherical geometries are less pronounced between the shock and the corotation radius than between the shock and the neutron star surface, but the difference of advection times are still significant in Fig.~\ref{fig:radv}.

The similar oscillation frequencies of the cylindrical and spherical equatorial geometries suggest that acoustic waves may play a dominant role when rotation is strong enough. We define the azimuthal acoustic radius $r_{\rm ac}$ such that the period $2\pi r_{\rm ac}/c$ of horizontal acoustic waves at this radius match the oscillation period $2\pi/\omega_r$. In the asymptotic regime of a large shock radius, the timescale associated to radial acoustic propagation is at least three times smaller than the azimuthal part because the perimeter of a circle is a factor $\pi$ larger than its diameter, and the adiabatic sound speed increases inward (like $c/c_{\rm sh}\sim (r_{\rm sh}/r)^{1/2}$) according to the Bernoulli equation.

\begin{figure}[t]
\centering
\includegraphics[width=\columnwidth]{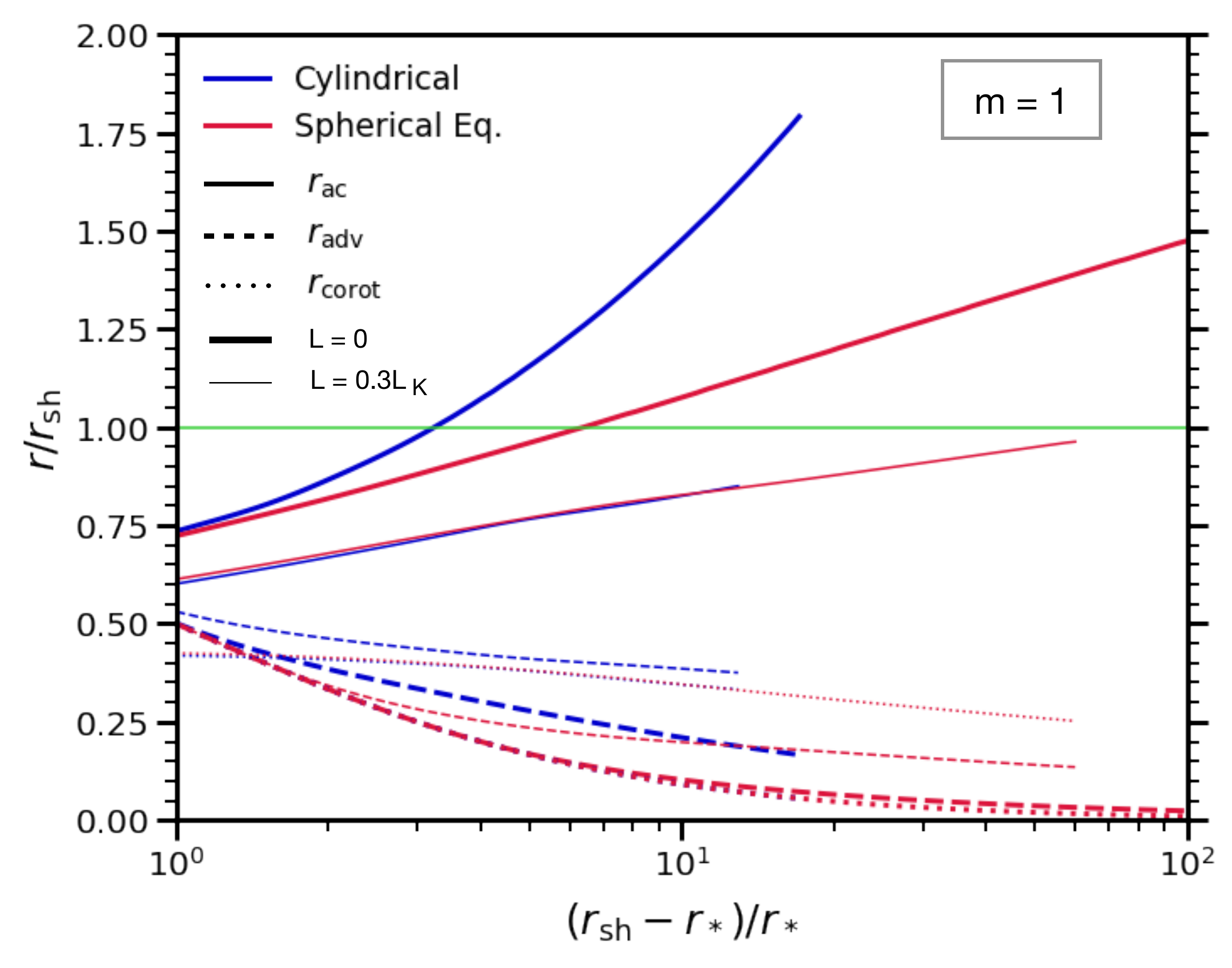}
\caption{Position of the corotation radius $r_{\rm corot}$ (dotted lines), advection radius $r_{\rm adv}$ (dotted lines), and  azimuthal acoustic radius $r_{\rm ac}$ (solid lines) for the mode $m=1$ in the cylindrical and spherical equatorial geometries (in blue and red, respectively). The values of $r_{\rm ac}$ exceeding $r_{\rm sh}$ are computed with a profile of the sound speed $c/c_{\rm sh}\sim (r_{\rm sh}/r)^{1/2}$ extrapolated above $r_{\rm sh}$ when necessary. The oscillation period is longer than any acoustic path without rotation (thick lines), whereas it coincides with an azimuthal acoustic path with rotation (thin lines).}
\label{fig:reff}
\end{figure}
We compare in Fig.~\ref{fig:reff} the values of $r_{\rm ac}$, $r_{\rm adv}$ and $r_{\rm corot}$ for the mode $m=1$ in cylindrical and spherical equatorial geometries, without rotation and with $L=0.3 L_{\rm K}$. Without rotation, the oscillation period is asymptotically longer than the azimuthal acoustic timescale at the shock, which is the longest acoustic path. However, with rotation, the oscillation period is always comparable to the azimuthal acoustic timescale in the vicinity of the shock, independently of the geometry of the flow.
The advection radius measured in Fig.~\ref{fig:reff} is asymptotically larger than the corotation radius in cylindrical geometry, and smaller than the corotation radius in spherical equatorial geometry, like in Fig.~\ref{fig:radv}.

These results suggest that the driving mechanism of spiral SASI in a rotating flow involves an acoustic resonance between horizontal acoustic waves and their radius of corotation. 
Figure~\ref{fig:reff} implies that the frequency of the unstable spiral mode exceeds the Lamb frequency at the shock (Eq.~\ref{defLamb}), allowing for a region of acoustic propagation immediately below the shock, irrespective of the flow geometry. 

The growth rate is more sensitive to the flow geometry than the oscillation frequency. 
It is possible that the corotational destabilization of horizontal acoustic waves contributes to this instability with an amplification mechanism based on the extraction of angular momentum from the inner region, which rotates faster than the wave pattern, and the outer region which rotates slower. The low frequency of this fundamental mode precludes a Wentzel-Kramers-Brillouin (WKB) approach, cf.~Ref.~\cite{Goldreich:1985} or Fig.~4 of Ref.~\cite{Yamasaki:2007dc}. 
We also note in Fig.~\ref{fig:OMEGA_PLOT} that the frequency of the dominant mode $m=2$ is larger the mode $m=1$ by a factor $\sim 1.4$ significantly smaller than the ratio $\sim 2$ of Lamb frequencies. Describing this instability mechanism as a simple resonance between horizontal acoustic waves at the shock and their corotation radius will require a more detailed analysis.

The challenges linked to the  use of  the WKB approach with low frequency modes had been an obstacle to elucidating the instability mechanism responsible for SASI in the absence of  rotation, when the advective-acoustic mechanism \cite{Foglizzo:2006fu,Foglizzo:2009,Guilet:2012} was advocated against the hypothesis of a purely acoustic mechanism \citep{Blondin:2006,Laming:2007}. We point out that our findings open the possibility of an acoustic mechanism only in the context of a rotating flow through the presence of a corotation radius.

It may also be possible that the pressure field generated by the advection of entropy and vorticity perturbations from the shock to the corotation radius mainly contributes to the growth rate of the instability, without affecting its frequency. 
These possible scenarios could be discriminated by considering an outer boundary condition which reflects acoustic waves without producing advected perturbations. This will be the focus of  future work.

\section{Conclusions}\label{sec:conclusions}

In order to test our understanding of the SASI mechanism in  non-rotating and   rotating stars, we have used a perturbative analysis of two flow geometries. 
We have focused on the shock dynamics in the equatorial plane and compared the properties of the instability in cylindrical and spherical equatorial geometries, in the asymptotic regime of  large shock radius.

We have established a unified set of differential equations valid for these two geometries, choosing variables such that the effect of rotation is well separated into a quadratic centrifugal force and a linear Doppler shifted frequency for non axisymmetric perturbation. The contribution of the azimuthal number $m$ describing non-axisymmetric perturbations is also well separated into its impact on the Lamb frequency for equatorial acoustic waves and the Doppler shifted frequency for rotating flows. This distinction allows to evaluate the accuracy of the spherical equatorial approximation compared to the full spherical geometry.
We find that the cylindrical and spherical models have very similar pressure and temperature structures and differ mainly by the advection time from the shock to the neutron star surface. We used this difference to test the impact of advection on the instability mechanism.

Relying on the results of Ref.~\cite{Foglizzo:2009}, we have predicted the efficiency of entropic-acoustic coupling in the region of strongest adiabatic compression near the neutron star surface. Using the properties of the stationary flow, we have shown that
phase mixing prevents an efficient entropic-acoustic coupling in this region in the cylindrical geometry, while it aids a strong entropic-acoustic cycle in the spherical equatorial geometry.

The important impact of the velocity profile on the efficiency of SASI suggests that it is also sensitive to the assumed value of the adiabatic index $\gamma=4/3$. Evaluating this sensitivity is however complex since $\gamma$ also affects the radius of the stalled shock, the Rankine-Hugoniot jump conditions, and the relative magnitude of entropy vs.~vorticity effects involved in 
the advective-acoustic coupling. 

Our numerical calculations of the eigenfrequencies with rotation in spherical equatorial geometry confirmed expectations as well as previous results obtained using numerical simulations. We also computed the propagation radius introduced in Ref.~\cite{Blondin:2017} and generalized its analytical dependence on the shock to neutron star radii ratio.

Surprisingly, we have found that  the cylindrical or spherical geometry of the equatorial flow has little impact on the oscillation frequency of the dominant spiral SASI in the presence of rotation. This strong contrast with the non-rotating configuration has led us to question the role of advection on the SASI mechanism in a rotating flow. We propose that the instability mechanism could be dominated by the exchange of energy and momentum across the corotation radius. 
A proper understanding of the effect of rotation on SASI should also account for the regime with modest rotation, too small to produce a corotation radius, but large enough to significantly impact  the efficiency of the SASI mechanism.  
Elucidating this mechanism will require a deeper investigation in future work.

\acknowledgments 
We thank A.C.~Buellet, M.~Bugli, and J.~Guilet  for helpful discussions.
This project has received funding from the  Villum Foundation (Project No.~37358), the Danmarks Frie Forskningsfonds (Project No.~8049-00038B), and the Deutsche Forschungsgemeinschaft through Sonderforschungbereich SFB 1258 ``Neutrinos and Dark 
Matter in Astro- and Particle Physics'' (NDM). LW thanks the Laboratoire AIM at CEA Saclay for hospitality during early stages of this work.

\appendix

\section{Differential system and boundary conditions}\label{sec:A4}

In the set of stationary equations, the geometrical factor $g$ mainly impacts the radial distribution of mass flux:
\begin{align}
    r^g \rho v_{r} &= r_{\rm sh}^g \rho_{\rm sh} v_{r_{\rm sh}}\ , \label{eq:mass_cons_ap}\\ 
    \frac{\partial S}{\partial r} &= \frac{\mathcal{L}}{P v_r}\ , \label{eq:entropy_cons_ap}\\
    \frac{\partial}{\partial r} \Bigg( \frac{v_r^2}{2}  + \frac{c^2}{\gamma - 1} + \Phi_{L^2} \Bigg) &= \frac{\mathcal{L}}{\rho v_r}\ ,\label{eq:euler_cons_ap}
\end{align}
with  $\rho = \gamma P/c^2$. We have incorporated the centrifugal effect into the modified potential $\Phi_{L^2}$ defined by
\begin{eqnarray}
\Phi_{L^2}&\equiv&\Phi_0+{L^2\over 2r^2}\ ,\\
{\partial \Phi_{L^2}\over\partial r}&=&{\partial \Phi_0\over\partial r}-{L^2\over r^3}\ .
\end{eqnarray}
The stability analysis is performed by integrating the differential system describing the stationary flow and perturbed quantities from the shock front to the neutron star surface marked by the radius $r_\ast$. 
At the inner boundary, the radial flow is steeply decelerated and abruptly halted. This introduces a mathematical singularity in the stationary flow equations. To overcome this challenge numerically, the variable $\log \mathcal{M}$ is adopted as integration variable when integrating the differential system from the shock down to the accretor, like in \cite{Foglizzo:2006fu}. The inner boundary is then defined as the point where the Mach number has reached a sufficiently small value  ($\mathcal{M} \sim 10^{-9}$). 
The perturbed flow equations are established using the same variables $(\delta v_\Phi,\delta h,\delta S,\delta K)$ defined by: 
\begin{eqnarray}
\label{eq:df_def}
\delta h &\equiv& \frac{\delta v_r}{v_r} + \frac{\delta \rho}{\rho} \ , \\
\delta S &\equiv& \frac{1}{\gamma-1} \frac{\delta c^2}{c^2} - \frac{\delta \rho}{\rho}\ , \\
{\delta K\over m^2} &\equiv&  {r v_r\over im} \delta w +  \frac{c^2}{\gamma}\delta S \ .\label{defdK}
\end{eqnarray}
In Eq.~(\ref{defdK}), the vertical vorticity perturbation $\delta w$ is measured along the vertical direction oriented downward, which coincides with $\delta w_\theta$ in spherical coordinates and $-\delta w_z$ in cylindrical coordinates. This definition of $\delta K$ is the same as Eq.~(C19) in \cite{Abdikamalov:2021} obtained in spherical equatorial geometry and coincides with the definition of $(im\delta K_1-im^2\delta q)$ in \cite{Yamasaki:2007dc} in cylindrical geometry.
Compared to the definition of $\delta K$ in previous studies in spherical geometry~\citep{Foglizzo:2001, Foglizzo:2006, Foglizzo:2006fu, Abdikamalov:2021}, the present definition of $\delta K$ involves $m^2$ rather than $l(l+1)$ because it is the eigenvalue of the angular part of the Laplacian operator in the equatorial plane, as already noted in cylindrical geometry by \cite{Yamasaki:2007dc} and in spherical equatorial geometry with rotation in \cite{Abdikamalov:2021}. The particular variables $(r\delta v_\Phi/im,\delta h,\delta S,\delta K/m^2)$ are chosen in order to obtain a differential system with the following properties:
\par(i) the frequency $\omega$ appears only as a doppler shifted frequency $\omega^\prime \equiv \omega-mL/r^2$;
\par(ii) apart from its linear effect on the doppler shifted frequency $\omega^\prime $, the specific angular momentum $L$ appears explicitly only as a quadratic centrifugal correction in $\Phi_{L^2}$; 
\par(iii) the geometrical parameter $g$ is explicitly absent except in the boundary conditions at the shock.

Of course the geometrical parameter $g$ is implicitly present through the radial profiles of density, pressure and more importantly the radial velocity of the stationary flow.
In comparison, the equivalent systems ($\delta f,\delta h, \delta S, \delta q$) defined in cylindrical geometry in \cite{Yamasaki:2007dc}, or ($\delta f,\delta h, \delta S, \delta K$) considered in spherical equatorial geometry in \cite{Abdikamalov:2021} do not satisfy the properties (i)--(iii).

The differential system describing the perturbed flow quantities in the post-shock layer is then
\begin{eqnarray}
{\partial \over \partial r}\left( {r\delta v_\phi\over im}\right)&=&
\delta v_r 
+{1\over v_r}\left(
{c^2\over\gamma}\delta S
-{\delta K\over m^2}
\right)\ ,\label{eq:dvphi_final}\\
\frac{\partial \delta h}{\partial r} &=& \frac{i \omega^\prime}{v_r} \frac{\delta \rho}{\rho} + \frac{m^2}{r^2v_r} {r\delta v_\phi\over im} \  , \label{eq:dh_final}\\
\Bigg( \frac{\partial}{\partial r} - \frac{i \omega^{\prime}}{v_r} \Bigg ) \delta S &=& \delta \Bigg( \frac{\mathcal{L}}{P v_r} \Bigg) \ ,  \label{eq:dS_final} \\
\Bigg(\frac{\partial}{\partial r} - \frac{i \omega^{\prime}}{v_r} \Bigg ) {\delta K\over m^2} &=& \delta \Bigg( \frac{\mathcal{L}}{\rho v_r} \Bigg) \ . \label{eq:dk_final}
\end{eqnarray}
The perturbed quantities in this differential system can be all expressed in terms of the four dynamical variables $(r\delta v_\Phi/im,\delta h,\delta S,\delta K/m^2)$ using the following relations:
\begin{eqnarray}
&&{\delta v_r\over v_r}={1\over 1-\mathcal{M}^2}\left(\delta h+\delta S-{i\omega^\prime \over c^2} {r\delta v_\phi\over im}-{\delta K\over m^2c^2} \right)\ ,\label{dvr}\\
&&{\delta \rho\over\rho}=\delta h -{\delta v_r\over v_r}\ ,\\
&&{\delta c^2\over c^2}=(\gamma-1)\left(\delta h +\delta S
-{\delta v_r\over v_r}\right)\ ,\\
&&\delta \Bigg( \frac{\mathcal{L}}{\rho v_r} \Bigg) = \nabla S \frac{c^2}{\gamma} \Bigg[ (\beta - 1) \delta h + \alpha \frac{\delta c^2}{c^2} -\beta \frac{\delta v_r}{v_r} \Bigg]\ , \\
&&\delta \Bigg( \frac{\mathcal{L}}{P v_r} \Bigg) = 
\nabla S  \Bigg[ (\beta - 1) \delta h + (\alpha-1) \frac{\delta c^2}{c^2} -\beta \frac{\delta v_r}{v_r} \Bigg]
\ .\label{eqdLPvr}
\end{eqnarray}
The explicit expressions for the coefficients 
 of the matrix ${\bf M_r}$ in Eq.~(\ref{matrix_form})
 are deduced from Eqs.~(\ref{eq:dvphi_final})--(\ref{eq:dk_final}) and Eqs.~(\ref{dvr})--(\ref{eqdLPvr}):
\begin{eqnarray}
M_r^{11}&\equiv&-{v_r\over 1-\mathcal{M}^2}{i\omega^\prime\over c^2},\label{Mr11}\\
M_r^{12}&\equiv&{v_r\over 1-\mathcal{M}^2}\ ,\\
M_r^{13}&\equiv&{c^2\over v_r}\left({\mathcal{M}^2\over 1-\mathcal{M}^2}+{1\over\gamma}\right)\ ,\\
M_r^{14}&\equiv&-{1\over v_r}{1\over 1-\mathcal{M}^2}\ ,\\
M_r^{21}&\equiv&-{1\over v_rc^2}{1\over 1-\mathcal{M}^2}
\left\lbrack\omega^{\prime2}-{m^2c^2\over r^2}(1-\mathcal{M}^2)\right\rbrack\ ,\label{Mr21}\\
M_r^{22}&\equiv&-{i\omega^\prime\over v_r}
{\mathcal{M}^2\over 1-\mathcal{M}^2}\ ,\\
M_r^{23}&\equiv&-{i\omega^\prime\over v_r}
{1\over 1-\mathcal{M}^2}\ ,\\
M_r^{24}&\equiv&{i\omega^\prime\over v_rc^2}
{1\over 1-\mathcal{M}^2}\ ,\\
M_r^{31}&\equiv&{\nabla S\over 1-\mathcal{M}^2}{i\omega^\prime\over c^2}
\left\lbrack
\beta+(\alpha-1)(\gamma-1)
\right\rbrack\ ,\\
M_r^{32}&\equiv&
-{\nabla S\over 1-\mathcal{M}^2}
\left\lbrace
1+\mathcal{M}^2\left\lbrack
\beta-1+(\alpha-1)  (\gamma-1)\right\rbrack \right\rbrace\ ,\\
M_r^{33}&\equiv&{i\omega^\prime\over v_r}
-{\nabla S\over 1-\mathcal{M}^2}
\left\lbrack\beta
+\mathcal{M}^2(\alpha-1)(\gamma-1)
\right\rbrack\ ,\\
M_r^{34}&\equiv&
{1\over c^2}{\nabla S\over 1-\mathcal{M}^2}
\left\lbrack\beta
+(\alpha-1)(\gamma-1)
\right\rbrack\ ,\\
M_r^{41}&\equiv&
{i\omega^\prime\over\gamma}{\nabla S\over 1-\mathcal{M}^2}
\left\lbrack
\beta+\alpha(\gamma-1)
\right\rbrack\ ,\\
M_r^{42}&\equiv&-{c^2\over\gamma}{\nabla S\over 1-\mathcal{M}^2}
\left\lbrace
1+\mathcal{M}^2\left\lbrack\beta-1+\alpha(\gamma-1)
\right\rbrack\right\rbrace\ ,\\
M_r^{43}&\equiv&
-{c^2\over\gamma}{\nabla S\over 1-\mathcal{M}^2}
\left\lbrack
\beta+\mathcal{M}^2\alpha(\gamma-1)
\right\rbrack\ ,\\
M_r^{44}&\equiv&{i\omega^\prime\over v_r}
-{\nabla S\over\gamma}{1\over 1-\mathcal{M}^2}
\left\lbrack
\beta+\alpha(\gamma-1)
\right\rbrack\ .\label{Mr44}
\end{eqnarray}
Compared to the differential system that would be obtained in spherical geometry without rotation \citep{Foglizzo:2006fu}, the only difference is the parameter $m^2$ replacing $l(l+1)$ in Eq.~(\ref{Mr21}), and the Doppler shifted frequency $\omega^\prime$ instead of $\omega$.

The boundary conditions required to numerically solve the linearized hydrodynamic equations are the same as in \cite{Foglizzo:2006fu}, modified by rotation as in \cite{Yamasaki:2007dc}. The conservation equations in the frame moving with the shock require to express the local gradients in the stationary flow:
\begin{eqnarray}
    \frac{\partial}{\partial r} \rho v_{r} = - \frac{g}{r} \rho v_r\ , \label{eq:stat1} \\
    \frac{\partial}{\partial r} \Bigg( P+\rho v_r^2 \Bigg) = 
    -\rho\left({{\rm d}\Phi_{L^2}\over {\rm d} r}
    +\frac{g}{r} v_r^2\right)  \ , \label{eq:stat2} \\
    \frac{\partial}{\partial r} \Bigg( \frac{v_r^2}{2}+ \frac{c^2}{\gamma - 1}+\Phi_{L^2} \Bigg) = \frac{\mathcal{L}}{\rho v_r} 
    \ .\label{eq:stat3}
\end{eqnarray}
where Eqs.~(\ref{eq:stat2}) and (\ref{eq:stat3}) correct some typos in Eqs.~(A43) and (A44) of \cite{Yamasaki:2007dc}.

The resulting boundary conditions are 
\begin{eqnarray}
\left({r\delta v_\phi\over im}\right)_{\rm sh} &=& 
\left(v_{r1}-v_{r2}\right)\Delta \zeta   \label{eq:BCvphi}\ ,\\
\label{eq:BCh}
\delta h_{\mathrm{sh}} &=&   
\left(1-{v_{r2}\over v_{r1}}\right)
\frac{\Delta v^{\prime}}{v_{\mathrm{sh}}}   \ ,\\
\nonumber
\delta S_\mathrm{sh} &=& 
{\Delta \zeta\over P_{\rm sh}}
\left\lbrack {\partial\over\partial r}(P+\rho v_r^2)
\right\rbrack^\mathrm{sh}_\mathrm{1}
-\Delta \zeta 
\left\lbrack\nabla S\right\rbrack^\mathrm{sh}_\mathrm{1}
\nonumber\\
&&
- 
\left(1-{v_{r2}\over v_{r1}}\right)^2
{\gamma v_{1r}v_{2r}\over c_{\rm sh}^2}
\frac{\Delta v^{\prime}}{v_\mathrm{sh}}
\label{eq:BCS0}\ ,\\
 &=& -\Delta \zeta 
(\nabla S_\mathrm{sh} -\nabla S_\mathrm{1})\nonumber\\
&&-\left(1-{v_{r2}\over v_{r1}}\right)
{\gamma \Delta \zeta\over c_{\rm sh}^2}
\left( {{\rm d}\Phi_{L^2}\over {\rm d} r}
-g{v_{r1}v_{r2}\over r_{\rm sh}}
\right) \nonumber\\
&&
- 
\left(1-{v_{r2}\over v_{r1}}\right)^2
{\gamma v_{1r}v_{2r}\over c_{\rm sh}^2}
\frac{\Delta v^{\prime}}{v_\mathrm{sh}}
\label{eq:BCS}\ ,\\
{\delta K_{\mathrm{sh}}\over m^2} &=& - \Delta \zeta \frac{c^2_{\mathrm{sh}}}{\gamma} 
\nabla S_{\mathrm{sh}}
\label{eq:BCk}
\ ,
\end{eqnarray} 
where $\Delta v^\prime\equiv -i\omega^\prime \Delta\zeta$. We note that the only impact of the geometrical parameter $g$ is in the expression of the entropy gradient, due to the change of slope of the momentum flux across the shock.

Under the assumption of a strong, adiabatic shock and assuming that the incoming gas is in free-fall ($v_{1r}^2+v_{1\phi}^2 = 2GM/r_{\rm sh}$) these boundary conditions are written as $\delta X_{\rm sh}=Y^{\rm sh}\Delta\zeta$ with the following components for the vector $Y^{\rm sh}$:
\begin{eqnarray}
Y^{\rm sh}_1 &\equiv& {2\over\gamma-1}v_{\rm sh}  \label{Ysh1}\ ,\\
\label{Ysh2}
Y^{\rm sh}_2 &\equiv&   
-\frac{2}{\gamma+1}
\frac{i\omega^{\prime}_{\rm sh}}{v_{\rm sh}}   \ ,\\
\label{Ysh3}\nonumber
Y^{\rm sh}_3 &\equiv& - \nabla S_\mathrm{sh} + \frac{\gamma}{\gamma+1} \frac{L^2}{c_\mathrm{sh}^2 r_\mathrm{sh}^3} - 
\\ &&\frac{1}{2 r_\mathrm{sh}}
\left({\gamma+1\over\gamma-1}-2g\right)
 + 
\frac{2}{\gamma + 1}
\frac{i\omega^{\prime}_{\rm sh}}{v_\mathrm{sh}}\ ,\\
Y^{\rm sh}_4 &\equiv& - 
\frac{c^2_{\mathrm{sh}}}{\gamma} 
\nabla S_{\mathrm{sh}}
\label{Ysh4}
\ .
\end{eqnarray} 
The only occurrences of the quadratic term $L^2$ and geometrical factor $g$ are in the expression of $Y^{\rm sh}_3$ defining the entropy perturbation at the shock $\delta S_{\rm sh}$.

At the inner boundary $r_{\rm NS}$, the requirement that the radial perturbation vanishes is deduced from Eq.~(\ref{dvr}):
\begin{eqnarray}
\left(\delta h+\delta S-{i\omega^\prime \over c^2} {r\delta v_\phi\over im}-{\delta K\over m^2c^2} \right)_{\ast}
=0\ ,\label{eq:BCNS}
\end{eqnarray}
which can be rewritten as $\delta X_{\ast}\cdot Y^{\ast}=0$ with the vector $Y^\ast$ defined by:
\begin{eqnarray}
Y^\ast_1&\equiv&-{i\omega^\prime_\ast \over c_\ast^2}\ ,\label{Yns1}\\
Y^\ast_2&\equiv&1\ ,\label{Yns2}\\
Y^\ast_3&\equiv&1\ ,\label{Yns3}\\
Y^\ast_4&\equiv&-{1\over c_\ast^2}\ .\label{Yns4}
\end{eqnarray}


\section{Amplification factor and  timescale of the SASI cycle}\label{sec:A_Q}

\begin{figure}[]
\centering
\includegraphics[width=\columnwidth]{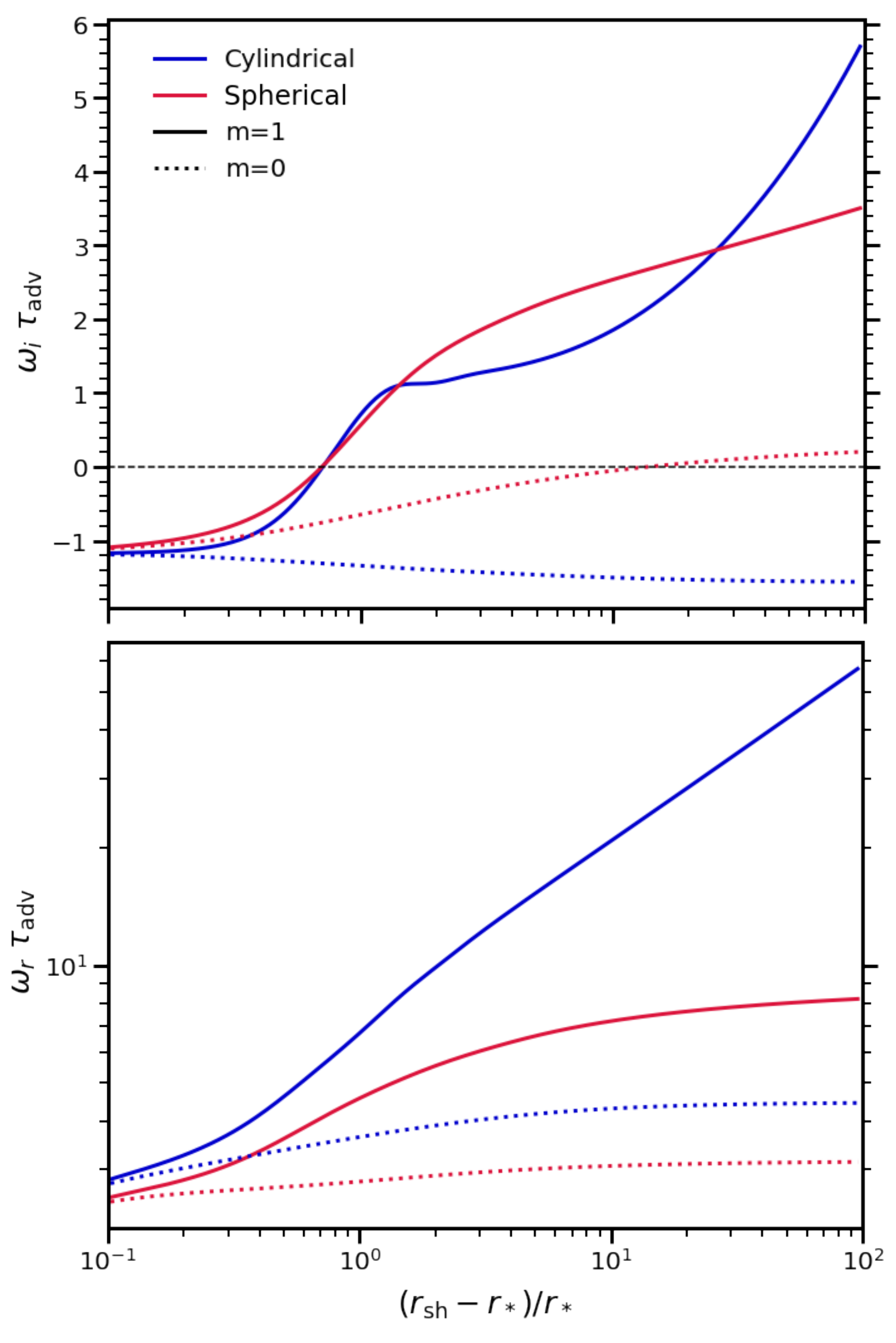}
\caption{Same as Fig.~\ref{fig:Eigenfrequency}, but using the advection time $\tau_{\rm adv}$ down to the neutron star surface to normalize the eigenfrequencies.}
\label{fig:Eigenfrequency_2}
\end{figure}
\begin{figure}[t]
\centering
\includegraphics[width=\columnwidth]{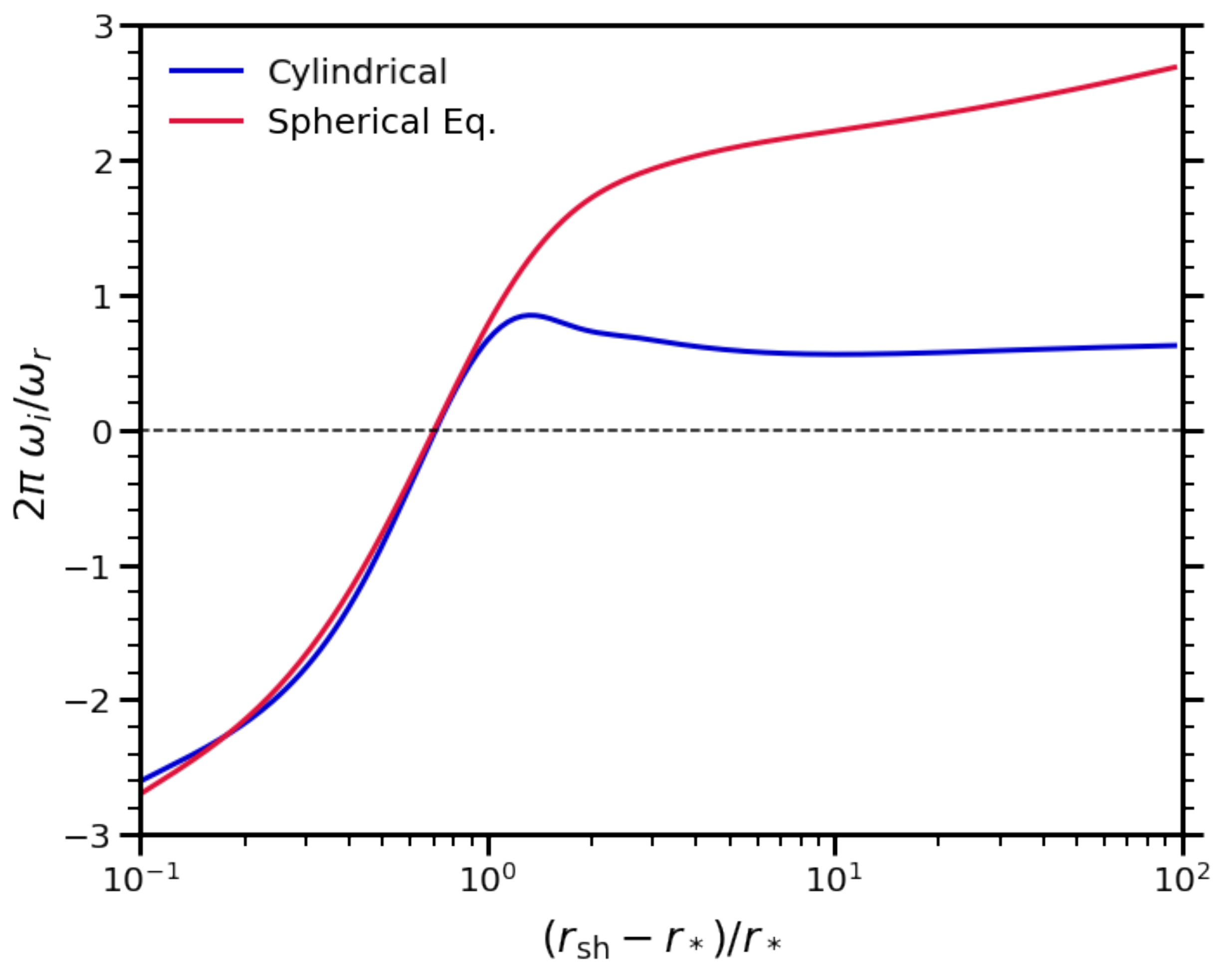}
\caption{Radial profile of the ratio $2 \pi {\omega_i}/{\omega_r}$ for the spherical and cylindrical models. For large shock radii this ratio becomes arbitrary large in the spherical case, while it is bounded in the cylindrical model. This implies a more prominent contribution of the entropic-acoustic coupling in the spherical system.}
\label{fig:Q}
\end{figure}
We illustrate in this Appendix the risk of misinterpretation associated to the choice of normalization of the eigenfrequencies. We reproduce in Fig.~\ref{fig:Eigenfrequency_2} the same eigenvalues as in Fig.~\ref{fig:Eigenfrequency}, but normalized with the advection time $\tau_{\rm adv}$ down to the neutron star surface rather than using $|v_{\rm sh}|/r_{\rm sh}$. This choice of normalization could be motivated by the fact that SASI is known to be driven by an advective-acoustic cycle, where perturbations are amplified by a complex factor ${\cal Q}\equiv|{\cal Q}|{\rm e}^{i\varphi_{\cal Q}}$ after each cycle of duration $\tau_{\cal Q}$. The additional contribution of the purely acoustic cycle was formalized in \cite{Foglizzo:2000} by the equation 
\begin{eqnarray}
{\cal Q}{\rm e}^{i\omega\tau_{\cal Q}}+{\cal R}{\rm e}^{i\omega\tau_{\cal R}}=1\ ,
\label{dispQR}
\end{eqnarray}
where $|{\cal R}|\le 1$ is the damping factor along a purely acoustic cycle of duration $\tau_{\cal R}$. 
In the regime $|{\cal Q}|\gg |{\cal R}|$ where the advective-acoustic cycle is sufficiently strong to make the purely acoustic contribution negligible, the 
the growth rate $\omega_i$ and oscillation frequency $\omega_r$ of the fundamental SASI mode associated to Eq.~(\ref{dispQR}) can be approximated by
\begin{eqnarray}
\omega_i&\sim& {1\over \tau_{\cal Q}}\log|{\cal Q}|\ ,\label{naive_wi}\\
\omega_r&\sim&{2\pi-\varphi_{\cal Q}\over \tau_{\cal Q}}\ .\label{naive_wr}
\end{eqnarray}
The magnitude of the offset of the oscillation frequencies with respect to the solution with $\varphi_{\cal Q}=0$ and $|{\cal R}|\ll |{\cal Q}|$ is illustrated in Figs. 2, 8 and 9 in \cite{Foglizzo:2009}. 

Estimating the advection time $\tau_{\cal Q}$ involved in SASI requires to identify the radial location of the region of most efficient advective-acoustic coupling, which is not necessarily on the neutron star surface.
If $\tau_{\rm adv}$ was a good approximation of $\tau_{\cal Q}$, $\omega_i\tau_{\rm adv}\sim \log |{\cal Q}|$ in Fig.~\ref{fig:Eigenfrequency_2} would directly measure the strength $|{\cal Q}|$ of the advective-acoustic cycle and would wrongly suggest that the amplification is stronger in the cylindrical geometry than in the spherical geometry. This interpretation is contradicted by the diverging behaviour of $\omega_r\tau_{\rm adv}$ in the bottom panel of Fig.~\ref{fig:Eigenfrequency_2}, whereas Eq.~(\ref{naive_wr}) would predict $\omega_r\tau_{\cal Q}\sim 2\pi-\varphi_{\cal Q}$. 

If Eqs.~(\ref{naive_wi})--(\ref{naive_wr}) were a good approximation, one could estimate the value of 
$\log |{\cal Q}|/(1-\varphi_{\rm Q}/2\pi)$ by measuring the ratio $2\pi\omega_i/\omega_r$ as in Fig.~\ref{fig:Q}. This figure indicates that this ratio can be large in the spherical model, but remains bounded $2\pi\omega_i/\omega_r\le0.6$ in the cylindrical model. The contribution of the purely acoustic cycle can be safely neglected when $|{\cal Q}|\gg1$ but can be important when $|{\cal Q}|$ is close to unity: Fig.~\ref{fig:Q} is thus consistent with a dominant advective-acoustic cycle in spherical geometry. 

We note that the asymptotically large value of $|{\cal Q}|$ may come from both contributions of entropic-acoustic and vortical-acoustic coupling. The identification of a region of strong adiabatic compression and the destabilization of the mode $m=0$ guarantee a major contribution of the entropic-acoustic coupling.

In the asymptotic regime where $|{\cal Q}|\gg1$ we can use $\omega_r$ and Eq.~(\ref{naive_wr}) to estimate the radius of most efficient advective-acoustic coupling if the phase shift $\varphi_{\cal Q}$ is negligible, as shown in Fig.~\ref{fig:radv}.

\bibliography{references.bib}

\end{document}